\documentclass[draft]{agujournal2019}
\usepackage{url} %this package should fix any errors with URLs in refs.
\usepackage{lineno}
\usepackage[inline]{trackchanges} %for better track changes. finalnew option will compile document with changes incorporated.
\usepackage{soul}
\usepackage{amsmath}
\usepackage{lscape}    % for 'landscape' environment
\usepackage{longtable} % for 'longtable' environment
\draftfalse

\journalname{ Journal of Advances in Modeling Earth Systems (JAMES)}

\begin{document}

%%%%%%%%%%%%%%%%%%%%%%%%%%%%%%%%%%%%%%%%%%%%%%%
%  TITLE
%
% (A title should be specific, informative, and brief. Use
% abbreviations only if they are defined in the abstract. Titles that
% start with general keywords then specific terms are optimized in
% searches)
%
%%%%%%%%%%%%%%%%%%%%%%%%%%%%%%%%%%%%%%%%%%%%%%%

% Example: \title{This is a test title}

\title{A simple model for the emergence of relaxation-oscillator convection}

%%%%%%%%%%%%%%%%%%%%%%%%%%%%%%%%%%%%%%%%%%%%%%%
%
%  AUTHORS AND AFFILIATIONS
%
%%%%%%%%%%%%%%%%%%%%%%%%%%%%%%%%%%%%%%%%%%%%%%%

% Authors are individuals who have significantly contributed to the
% research and preparation of the article. Group authors are allowed, if
% each author in the group is separately identified in an appendix.)

% List authors by first name or initial followed by last name and
% separated by commas. Use \affil{} to number affiliations, and
% \thanks{} for author notes.
% Additional author notes should be indicated with \thanks{} (for
% example, for current addresses).

% Example: \authors{A. B. Author\affil{1}\thanks{Current address, Antartica}, B. C. Author\affil{2,3}, and D. E.
% Author\affil{3,4}\thanks{Also funded by Monsanto.}}

\authors{F. E. Spaulding-Astudillo\affil{1} and J. L. Mitchell\affil{1,2}}

% \affiliation{1}{First Affiliation}
% \affiliation{2}{Second Affiliation}
% \affiliation{3}{Third Affiliation}
% \affiliation{4}{Fourth Affiliation}

\affiliation{1}{Department of Earth, Planetary, and Space Sciences, University of California, Los Angeles\newline595 Charles E Young Dr E, Los Angeles, CA 90095, USA}
\affiliation{1}{Department of Atmospheric and Oceanic Sciences, University of California, Los Angeles\newline520 Portola Plaza, Los Angeles, CA 90095, USA}
%(repeat as many times as is necessary)

% Corresponding author mailing address and e-mail address:

% (include name and email addresses of the corresponding author.  More
% than one corresponding author is allowed in this LaTeX file and for
% publication; but only one corresponding author is allowed in our
% editorial system.)

% Example: \correspondingauthor{First and Last Name}{email@address.edu}

\correspondingauthor{Francisco E. Spaulding-Astudillo}{fspauldinga@ucla.edu}

%%%%%%%%%%%%%%%%%%%%%%%%%%%%%%%%%%%%%%%%%%%%%%%
% KEY POINTS
%%%%%%%%%%%%%%%%%%%%%%%%%%%%%%%%%%%%%%%%%%%%%%%
%  List up to three key points (at least one is required)
%  Key Points summarize the main points and conclusions of the article
%  Each must be 140 characters or fewer with no special characters or punctuation and must be complete sentences

% Example:
% \begin{keypoints}
% \item	List up to three key points (at least one is required)
% \item	Key Points summarize the main points and conclusions of the article
% \item	Each must be 140 characters or fewer with no special characters or punctuation and must be complete sentences
% \end{keypoints}

\begin{keypoints}
\item A transition from quasi-equilibrium (QE) to relaxation-oscillator (RO) convection occurs at high surface temperatures on Earth.
\item QE breakdown is predicted by a simple model of a convective heat engine in radiative-convective equilibrium where plumes have zero buoyancy.
\item QE breaks down when the equilibrium condition of the heat engine is violated, and this leads to RO emergence.
\end{keypoints}

%%%%%%%%%%%%%%%%%%%%%%%%%%%%%%%%%%%%%%%%%%%%%%%
%
%  ABSTRACT and PLAIN LANGUAGE SUMMARY
%
% A good Abstract will begin with a short description of the problem
% being addressed, briefly describe the new data or analyses, then
% briefly states the main conclusion(s) and how they are supported and
% uncertainties.

% The Plain Language Summary should be written for a broad audience,
% including journalists and the science-interested public, that will not have 
% a background in your field.
%
% A Plain Language Summary is required in GRL, JGR: Planets, JGR: Biogeosciences,
% JGR: Oceans, G-Cubed, Reviews of Geophysics, and JAMES.
% see http://sharingscience.agu.org/creating-plain-language-summary/)
%
%%%%%%%%%%%%%%%%%%%%%%%%%%%%%%%%%%%%%%%%%%%%%%%

%% \begin{abstract} starts the second page

\begin{abstract}
Earth's tropics are characterized by quasi-steady precipitation with small oscillations about a mean value, which has led to the hypothesis that moist convection is in a state of quasi-equilibrium (QE). In contrast, very warm simulations of Earth's tropical convection are characterized by relaxation-oscillator-like (RO) precipitation, with short-lived convective storms and torrential rainfall forming and dissipating at regular intervals with little to no precipitation in between. We develop a model of moist convection by combining a zero-buoyancy model of bulk-plume convection with a QE heat engine model, and we use it to show that QE is violated at high surface temperatures.  We hypothesize that the RO state emerges when the equilibrium condition of the convective heat engine is violated, i.e., when the heating rate times a thermodynamic efficiency exceeds the rate at which work can be performed. We test our hypothesis against one- and three-dimensional numerical simulations and find that it accurately predicts the onset of RO convection. The proposed mechanism for RO emergence from QE breakdown is agnostic of the condensable, and can be applied to any planetary atmosphere undergoing moist convection. To date, RO states have only been demonstrated in three-dimensional convection-resolving simulations, which has made it seem that the physics of the RO state requires simulations that can explicitly resolve the three-dimensional interaction of cloudy plumes and their environment. We demonstrate that RO states also exist in single-column simulations of radiative-convective equilibrium with parameterized convection, albeit in a different surface temperature range and with much longer storm-free intervals.
\end{abstract}

\section*{Plain Language Summary}
Earth's tropics are characterized by steady rainfall, indicating that moist convection is a continuous process. However, in simulations of very warm conditions, a form of episodic convection emerges that is characterized by short bursts of intense rainfall followed by longer rain-free intervals. We construct a simple model that represents convection as a heat engine, and use it to show that steady convection must break down in very warm conditions. We hypothesize that the essential condition for steady convection is the balanced conversion of heat into work, which is violated at high surface temperatures. We test our hypothesis against climate model simulations of increasing complexity - the first parameterizes convection and the second actually resolves it - and found that it accurately predicts when the steady climate state transitions to the episodic state. The simple model of convection isn't limited to Earth, and could be applied to planets with different atmospheric compositions. Finally, while it has seemed that episodic precipitation could only be obtained from model simulations that resolve convection, we've shown here that it can also occur in simpler climate models with parameterized convection.

%%%%%%%%%%%%%%%%%%%%%%%%%%%%%%%%%%%%%%%%%%%%%%%
%
%  BODY TEXT
%
%%%%%%%%%%%%%%%%%%%%%%%%%%%%%%%%%%%%%%%%%%%%%%%

%%% Suggested section heads:
% \section{Introduction}
%
% The main text should start with an introduction. Except for short
% manuscripts (such as comments and replies), the text should be divided
% into sections, each with its own heading.

% Headings should be sentence fragments and do not begin with a
% lowercase letter or number. Examples of good headings are:

% \section{Materials and Methods}
% Here is text on Materials and Methods.
%
% \subsection{A descriptive heading about methods}
% More about Methods.
%
% \section{Data} (Or section title might be a descriptive heading about data)
%
% \section{Results} (Or section title might be a descriptive heading about the
% results)
%
% \section{Conclusions}

\section{Introduction}\label{sec:intro}

%\begin{figure*}[ht!]
\begin{figure*}
\centering
\noindent \includegraphics[width=\textwidth,angle=0]{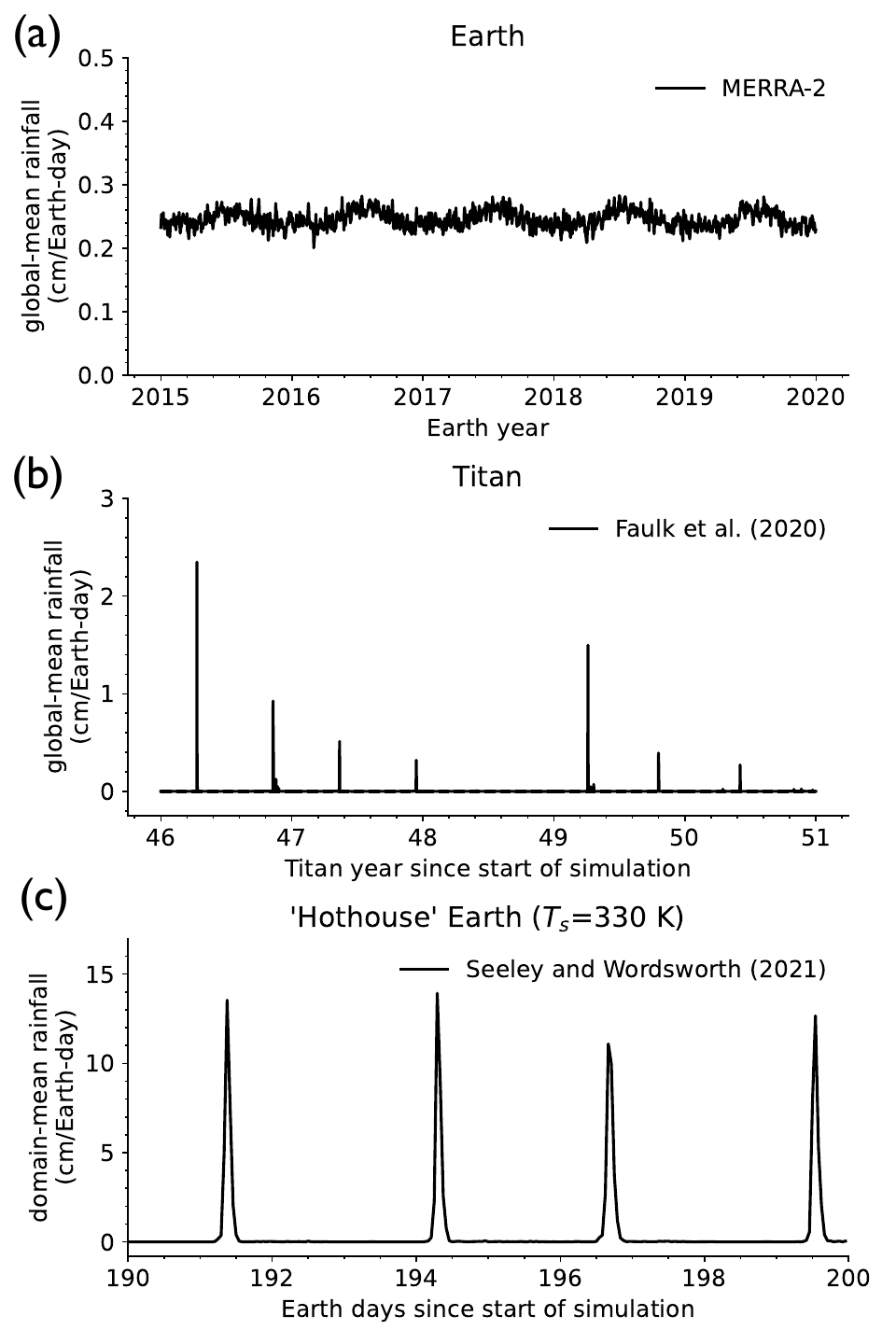}
\caption{Comparison of global-mean precipitation on (a) Earth from the MERRA-2 Earth climate reanalysis \cite{Gelaro2017} and (b) Titan from Titan Atmospheric Model simulations with best-fit land surface hydrology from \citeA{Faulk2020}, respectively. In (c), we show the domain-averaged precipitation in convection-resolving model aquaplanet simulations of the ``Hothouse Earth" at a fixed surface temperature of 330 K from \citeA{Seeley2021}.} \label{fig:precip-earth-titan}
\end{figure*}

To investigate the nature of convective dynamics in moist planetary atmospheres under varying conditions, we often rely on numerical climate models. These models are differentiated by whether or not they resolve convective processes, which refers to vertical motion at the scale of individual clouds (``plumes"). Models that can resolve convection are known to simulate the climate more realistically, but can be computationally expensive to run over very large domains. In coarse-resolution climate models that simulate the three-dimensional structure of the atmosphere on regional or global scales, parameterizations are employed to represent the net (``bulk") effect of an ensemble of unresolved plumes that turbulently mix with the surrounding environment. The bulk-plume representation of convection relates the properties of the plume ensemble (consisting of numerous members of varying size and intensity) to the large-scale radiative forcing through the concept of quasi-equilibrium \cite<QE;>[]{Arakawa1974,Emanuel2001}. QE is, by definition, a \textit{hypothesis} that there exists a steady balance between generation and dissipation of kinetic energy in convecting atmospheres \cite{Yano2012}. The QE hypothesis has been invoked successfully in various contexts, for example, to reveal the dynamics of mature hurricanes \cite{Emanuel1986,Emanuel2001}, formulate scaling laws for the maximum intensity of convective storms \cite{Emanuel1996}, derive analytical solutions for the large-scale tropical circulation \cite{Neelin2000}, and resolve the closure problem in the bulk-plume parameterization of convection \cite{Arakawa1974}. Thus, while the precise range of states over which QE is valid remains unclear \cite{Yano2012}, its utility in advancing conceptual understanding of moist convection is evident. 

In the contemporary solar system, there are two planets with similar atmospheric compositions but distinct convective dynamics: Earth and Saturn's moon Titan. Their atmospheres are nitrogen-dominated, and each has a condensing component that participates in an important ``hydrological" cycle, i.e.~the condensable undergoes phase changes to form clouds and precipitation \cite{Horst2017}. Whereas the condensing component on Earth is obviously water, the atmosphere of Titan is so cold that methane (and to a lesser extent nitrogen) condenses \cite{Mitchell2016,Tokano2017} while water is part of the icy bedrock of the Titan's surface \cite{Griffith2003}. The pattern of global-mean precipitation on Earth (Figure \ref{fig:precip-earth-titan}a) from the MERRA-2 \cite{Gelaro2017} global climate reanalysis and Titan (Figure \ref{fig:precip-earth-titan}b) from global simulations of Titan by \citeA{Faulk2020} reveals important differences in their hydrological cycles. On Earth, global-mean precipitation is quasi-steady about a mean value with small oscillations (Figure \ref{fig:precip-earth-titan}a), suggesting that modern Earth has QE-type convection.  On Titan, however, rain is the exception, rather than the rule (Figure \ref{fig:precip-earth-titan}b); storms erupt at semi-regular intervals with vigorous, short-lived rainfall of several centimeters per Earth day and extended dry spells in between \cite{Battalio2022}. The largest observed storms produce cloud cover up to 10\% of Titan's disk, in contrast to most observations showing very little cloud cover \cite{Griffith1998, Schaller2009}. During these storms \cite{Dhingra2019,Turtle2011b}, Titan's surface is likely subject to fluvial erosion which carves out channels and valleys and discharges the sediments into alluvial fans \cite{Perron2006,Horst2017,Faulk2017,Lewis-Merrill2022}. 

Where does this striking difference in surface precipitation originate? In this study, we take the first step towards answering that question by taking a holistic view of the role of moisture in radiative-convective processes. For completeness, we begin with some background on the physics of moist convection as it relates to convective storms. The temperature of a moist air parcel decreases less rapidly with height than a dry air parcel displaced from the same initial location because of latent heat release. If the moist parcel is displaced adiabatically (i.e., without exchanging heat or mass with its surroundings) above its saturation level, then it usually becomes warmer than the sub-saturated environment in which it is embedded, and therefore accelerates upward due to buoyancy, condensing moisture along its path. The low molecular weight of water vapor on Earth and methane vapor on Titan relative to the dry background gases \cite{Seidel2020,Mitchell2016} lends additional buoyancy to moist parcels of air \cite{DaYang2020}. The vertical integral of buoyancy $B$ (m/s$^2$) along the upward trajectory of an adiabatic parcel can be decomposed into the meteorological quantities known as convective available potential energy (CAPE; J kg$^{-1}$) and convective inhibition (CIN; J kg$^{-1}$):
\begin{equation}
    \label{eq:Bdz_net}
    \int_{0}^{\text{LNB}} B \: dz = \text{CAPE} - \text{CIN},
\end{equation}
where
\begin{equation}
\label{eq:cape(rho)}
\text{CAPE} = \int_{\text{LFC}}^{\text{LNB}} B \: dz \: \: \text{and} \: \: \text{CIN} = -\int_{0}^{\text{LFC}} B \: dz
\end{equation}
are both expressed as positive values, LFC is the level of free convection, and LNB is the level of neutral buoyancy. Equation \ref{eq:Bdz_net} represents the net work done to lift the adiabatic parcel from the surface to the LNB, and can be positive or negative. CAPE is an important meteorological quantity because it represents the maximum intensity of convective storms and is, for example, correlated with the frequency distribution of lightning flashes on Earth \cite{Romps2018}. CIN can be thought of as the energetic barrier to convection. In the global-mean, $\text{CAPE}-\text{CIN}>0$ on Earth \cite{Riemann2009} and there is some evidence to suggest that $\text{CAPE}-\text{CIN}<0$ on Titan \cite{Battalio2022}. Given our interest in the maximum possible amount of work done by the climate system, we approximate the net vertical integral of buoyancy (Equation \ref{eq:Bdz_net}) as CAPE for the remainder of this study. Neglecting CIN could be a poor assumption in, for example, hydrogen atmospheres where the vapor phase of the condensing substance is significantly heavier than the dry background gas. CAPE measures the positive buoyancy that is generated by the absorption of solar radiation at the surface and emission of planetary radiation to space in the troposphere, and represents the ``fuel" for thunderstorms. Coincidentally, regions of Titan with elevated near-surface humidity ($\sim$60\%) have similar values of CAPE to the modern-day tropics of Earth \cite{Riemann2009,Griffith2008,Tokano2006,Barth2007,Seeley2023}, indicating the potential for intense storms and rainfall. 

It has recently come to light that there is an emergent \textit{dynamical similarity} between contemporary Titan and a hotter Earth (Figure \ref{fig:precip-earth-titan}b,c). The key discovery, in this case, was made by \citeA{Seeley2021} in a study of Earth's tropical clouds and precipitation, in which the authors incrementally increased the surface temperature in a convection-resolving model. At surface temperatures above 320 K, they discovered a ``hothouse" climate state with a new mode of convection undergoing relaxation oscillations. The convective oscillations produced deluges lasting a few hours that then repeated every few days (Figure \ref{fig:precip-earth-titan}c). Unlike modern Earth, hothouse (often referred to as ``moist greenhouse") climate simulations have radiative heating in the lower troposphere \cite{W&T}.  \citeA{Seeley2021} hypothesized this lower-tropospheric radiative heating (LTRH) is a necessary condition for the RO state.  LTRH occurs in hothouse climates because of the thermodynamic and radiative properties of water vapor. Around 320 K, the ``water vapor window" - a spectral region over which the present-day atmosphere is transparent to infrared radiation - closes \cite{Pierrehumbert2010,Koll2018}, which prevents the lower atmosphere from directly cooling to space \cite{W&T}. \citeA{Seeley2021} tested the LTRH hypothesis by carrying out a series of experiments with fixed radiative cooling profiles with and without LTRH. In cases with imposed LTRH, RO states emerged at much lower temperatures close to the modern-day tropics ($\sim$290 K). In cases without LTRH, no RO states emerged. Two subsequent studies found that the RO state can emerge in the absence of LTRH \cite{Dagan2023,Song2024}. Given the available evidence, LTRH is not a necessary condition for RO emergence, though it can support it.

There is a consensus among the aforementioned studies that water vapor plays an important role in the emergence of RO convection at high temperatures on Earth. In what follows, we explore the nature of that role. In QE convection, kinetic energy is generated and dissipated in the atmosphere at equal rates, conceivably leading to steady precipitation.  Since the RO state is, by definition, non-steady, it suggests that we may conceive of the QE-to-RO transition as a \textit{breakdown of quasi-equilibrium}. To look for a breakdown of QE convection with increasing surface temperature (and/or, as we will see, moisture content), what is needed is a plausible model of QE convection. Since the atmosphere is a compressible system, parcels/plumes exchange energy with the environment through heat exchange and work, the simplest QE model of convection is that of a heat engine \cite{Renno1996,Emanuel1996} defined in the traditional way as any closed system that converts heat into work at some thermodynamic efficiency. In the context of an atmosphere, the work done by the convective heat engine is known as CAPE and heat is transferred to and from the system in the form of radiation \cite{Arakawa1974,Emanuel1986}. 

The dynamical similarity of Titan and the hothouse Earth (Figure \ref{fig:precip-earth-titan}a,c) could point to an underlying physical mechanism that is general to both planets, and their example motivates us to search for an explanation that is inclusive of both radiative and convective processes. A heat engine theory of convection would be agnostic of the atmospheric composition and the condensing substance. For this reason, it is an ideal framework to compare the atmospheres of Earth and Titan. While a goal for future work is to establish whether the theory can explain Titan’s bursty methane precipitation (Figure \ref{fig:precip-earth-titan}b), the specific aim of this study is to apply the theory to Earth.  

\section{Theory}\label{sec:theory}
The QE state is characterized by a steady balance between the generation of CAPE by radiation and its conversion into kinetic energy - i.e., convective motion. The RO state is clearly not steady, however we hypothesize that exploring the conditions in which the steady, QE state is valid can illuminate the mechanisms that lead to the transition from QE to RO convection. First, we construct a quasi-equilibrium model of a convective heat engine \cite<Section \ref{sec:HE};>[]{Emanuel1996} in which entraining clouds have zero-buoyancy relative to their local environment \cite<Section \ref{sec:ZBA};>[]{S&O,Romps2016}. Second, we demonstrate that QE-type convection is inconsistent with the energetic requirements of radiative-convective equilibrium when the surface is sufficiently warm and/or humid (Section \ref{sec:condition}). Thus, we predict that RO-type convection emerges in warm and humid conditions. A detailed list of the symbols, constants, and acronyms introduced in this section is given in \ref{app:A}.

\begin{figure*}
  \centering
  \includegraphics[width=\textwidth,angle=0]{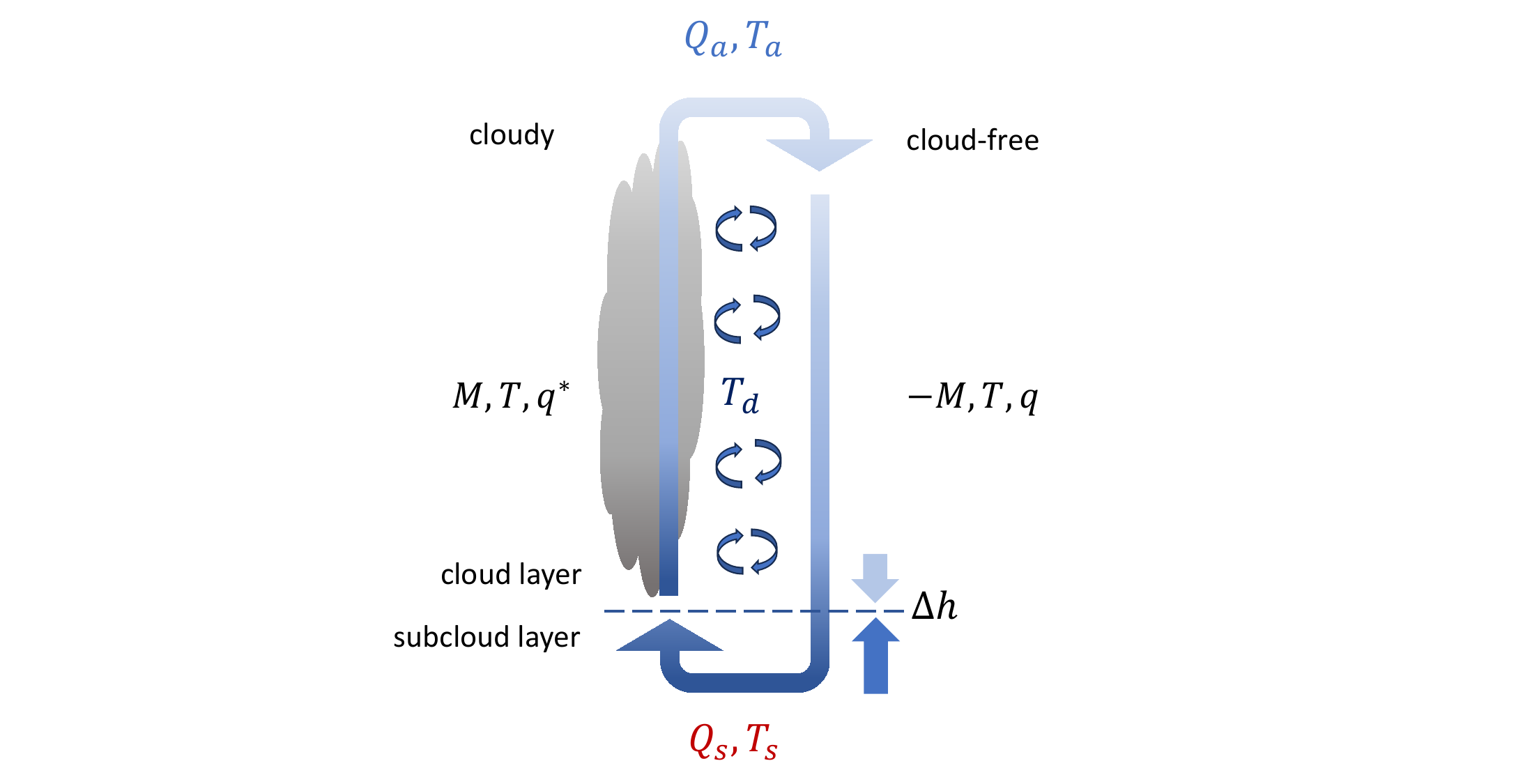}
  \caption{Schematic illustration of the zero-buoyancy heat engine model of convection based on \citeA{Emanuel1996} and \citeA{Romps2016}. Radiation is absorbed by the surface at warm temperature $T_s$ at a rate $Q_s$ and emitted by the atmosphere at the cold temperature $T_a$ at a rate $Q_a$. The mass flux $M$ is assumed to be constant with height and equal in magnitude in ascending cloudy and descending cloud-free regions, which are permitted to turbulently mix. The rate of work is $M \times \textrm{CAPE}$. Heat is converted into work at a thermodynamic efficiency $\eta$ that is a ratio of the temperatures at which irreversible frictional dissipation occurs ($T_d$) and radiation is absorbed and emitted. In the zero-buoyancy approximation, cloudy regions and cloud-free regions have the same temperature $T$ but different specific humidities at each height. In statistical equilibrium, mixing equal parts of air between the cloud layer and subcloud layer implies a net upward transport of moist static energy ($\Delta h$).} \label{fig:HE}
\end{figure*}

\subsection{Theory 1: convection as a heat engine}\label{sec:HE}

Here, we summarize a quasi-equilibrium model of convection developed by \citeA{Emanuel1996} and others \cite{Renno1996,Pauluis2002a}. We begin with the entropy budget of the climate system (defined as the surface and the atmosphere) in statistical equilibrium as commonly written \cite<e.g.,>[]{Pauluis2002a,Singh2022a}:
\begin{equation}
    \label{eq:Sbudget}
    \dot S_{rad} + \dot S_{irr} = 0,
\end{equation}
where $\dot S_{rad}<0$ is the entropy change associated with radiative processes and $\dot S_{irr}>0$ is the entropy change associated with irreversible processes within the climate system. By convention, time rates of change of entropy have units of W K$^{-1}$. Since Earth receives low entropy energy from the sun and releases high entropy energy to space, radiative processes represent a sink of entropy \cite{Singh2022a}. Dry processes such as dissipation of convective turbulence and moist processes such as hydrometeor sedimentation, phase changes, and vapor diffusion irreversibly increase the entropy of the climate system \cite{Pauluis2002a}. In a dry atmosphere, the dominant contributor to entropy generation in $\dot S_{irr}$ is frictional dissipation of convective turbulence \cite{Singh2022a}. Although Earth's atmosphere is not dry, frictional processes remain a major source of dissipation because of hydrometeor drag \cite{Pauluis2002a,Singh2016}. In what follows, we assume that the dominant source of entropy is frictional dissipation ($\dot S_d$; W K$^{-1}$), which should be understood as referring to both convective turbulence and hydrometeor drag. This key assumption of the heat engine model simplifies the entropy budget and will lead us to a solution for the equilibrium mass flux. The relative importance of frictional dissipation in the entropy budget over a range of surface temperatures was recently explored \cite{Singh2016}, where it was shown to account for approximately 50\% of the total entropy generation at 305 K and increase with further surface warming. Some caveats of our approach are addressed in Section \ref{sec:dis} and at greater length by several preceding studies \cite{Singh2022a,Pauluis2002a,Emanuel1986}. The spatially-resolved flux of radiation at the surface is $\tilde{F}_s(x,y)$ (W m$^{-2}$), so the rate of energy transfer to the surface by radiation (W) is
\begin{equation}
    Q_s = \int_A \tilde F_s,
\end{equation}
where $\int_A  = \int dx \: dy$ denotes an integral over the horizontal area of the system. Similarly, the spatially-resolved atmospheric cooling rate per unit mass is $ \tilde Q_a(x,y,z)$ (W kg$^{-1}$). The rate of energy loss from the atmosphere by radiation (W) is
\begin{equation}
    \label{eq:Qs=QA}
    Q_a = \int_V \rho \: \tilde Q_a,
\end{equation}
where $p$ is pressure, $\rho$ is density, $g$ is gravity, and $\int_V  = \int dx \: dy \: dz = -\int_A \int \frac{dp}{\rho g}$ is an integral over the system volume. Atmospheric cooling ($F_a$; W m$^{-2}$) balances surface heating ($F_s$; W m$^{-2}$) in equilibrium:
\begin{equation}
    \label{eq:Ebudget}
     F_s +  F_a = \frac{Q_s}{A} + \frac{Q_a}{A} = 0,
\end{equation}
where $A$ is the horizontal area of the system. Radiative processes are assumed to be thermodynamically reversible where absorption and emission occur at a mean temperature of $T_s$ and $T_a$. The spatially-resolved rate of change of specific entropy associated with radiative processes (W kg$^{-1} $K$^{-1}$) is 
\begin{equation}
    \dot s_{rad} = \frac{\tilde Q_a }{T}.
\end{equation}
The total rate of decrease in entropy of the system ($\dot S_{rad}$; W K$^{-1}$) is
\begin{equation}
    \label{eq:Srad}
    \dot S_{rad} = \int_V \rho \dot s_{rad} = \frac{Q_a}{T_a} + \frac{Q_s}{T_s} = Q_a \Bigg({1\over T_a}-{1\over T_s}\Bigg),
\end{equation}
where 
\begin{equation}
    \label{eq:Trad}
    \frac{1}{T_a} = \frac{\int_V \rho \tilde Q_a/T}{\int_V \rho \tilde Q_a}.
\end{equation}
Invoking quasi-equilibrium, we assume that the rates of kinetic energy generation ($\dot W$; W) and dissipation ($\dot D$; W) are equal. By definition, $\dot W$ is also the rate of work. It can be shown that, in statistical equilibrium, these processes are balanced in the volume integral of the ``kinetic energy equation" \cite<i.e., the equation describing the time rate of change of kinetic energy per unit volume;>[]{Emanuel1996,Pauluis2002a}:
\begin{equation}
    \label{eq:KE}
    \dot W - \dot D = 0,
\end{equation}
where 
\begin{equation}
    \dot W = \int_V -v \cdot \nabla p \: \: \textrm{and} \: \: \dot D = -\int_V\rho f \cdot v. \ 
\end{equation}
Here, $v$ is the velocity and $f$ is the frictional force per unit mass. The spatially-resolved rate of change of specific entropy associated with frictional dissipation (W kg$^{-1}$ K$^{-1}$) is
\begin{equation}
    \dot s_d = -\frac{f \cdot v}{T}.
\end{equation}
Frictional dissipation is an irreversible source of entropy ($\dot S_d$; WK$^{-1}$); assuming it occurs at a mean temperature $T_d$,
\begin{equation}
    \label{eq:Sdis}
    \dot S_d = \int_V \rho \dot s_d  = - {1 \over T_d} \int_V \rho f \cdot v = \frac{\dot D}{T_d} = \frac{\dot W}{T_d}.
\end{equation}
Following \citeA{Pauluis2002a}, we can now re-write the entropy budget (Equation \ref{eq:Sbudget}) first by defining a new variable $\Delta \dot S = \dot S_{irr} - \dot S_d$ to be all irreversible sources of entropy \textit{except} frictional dissipation and second by substituting $\dot S_{rad}$ (Equation \ref{eq:Srad}) and $\dot S_d$ (Equation \ref{eq:Sdis}):
 
\begin{equation}
    \label{eq:Sbudget-with-dS}
    Q_a \Bigg({1\over T_a}-{1\over T_s}\Bigg) + \frac{\dot W}{T_d} + \Delta \dot S = 0.
\end{equation}
Re-arranging this expression and solving for the rate of work, we find
\begin{equation}
    \label{eq:Wmax}
    \dot W = - Q_a T_d \Bigg({1\over T_a}-{1\over T_s}\Bigg) - T_d \Delta \dot S \approx -\eta Q_a,
\end{equation}
where 
\begin{equation}
    \label{eq:eta}
    \eta = T_d \Big(\frac{1}{T_a} - \frac{1}{T_s} \Big)
\end{equation}
is the efficiency of the heat engine when frictional dissipation is the only source of entropy generation. Equation \ref{eq:Wmax} nicely demonstrates that the other sources of entropy $\Delta \dot S$ reduce the amount of work that can be performed by the heat engine. As a logical starting place, in the second step of Equation \ref{eq:Wmax}, we assume that $\Delta \dot S=0$. Indeed, \citeA{Pauluis2002a} point out that, in making this assumption, one is actually solving for the maximum rate of work that can be performed given the typical forcing on and state of the system. \citeA{Emanuel1996} show that to a good approximation
\begin{equation}
    \label{eq:W_MB}
    \dot W \approx \int_V MB,
\end{equation}
where $M$ is the mass flux (kg m$^{-2}$ s$^{-1}$) and 
\begin{eqnarray}
    \label{eq:B}
    B = g {(\overline \rho - \rho) \over \rho}
\end{eqnarray}
is the buoyancy (m s$^{-2}$), $\overline \rho$ is the mean density of the system, and $\rho$ is the spatially-resolved density of the working fluid (i.e.~the parcel/plume); $MB$ is the buoyancy flux (W m$^{-3}$). The above imply the rate of work is equal to the buoyancy flux. We can simplify Equation \ref{eq:W_MB} as follows:
\begin{align}
\int_V MB &= \int_z \int_{A_u} M_u B_u + \int_z \int_{A_d} M_d B_d  \nonumber \\
&\approx \: A_u M_u \int_z B_u + A_d M_d \: \int_z B_d \nonumber \\
\label{eq:MB_simple}
&\approx  A \times |M| \times \textrm{CAPE}
\end{align}
where $A_u$, $A_d$ and $M_u$, $M_d$ are the respective areas and mass fluxes of the updrafts and downdrafts. Together, they span the system area: $A_u + A_d = A$. In the first step, we decompose the volumetric integral over the system into updrafts and downdrafts. In the second step, we assume that the mass fluxes are constant (true, for example, in an adiabatic plume) and that buoyancy of updrafts and downdrafts is horizontally homogeneous. Buoyancy is approximately horizontally homogeneous, for instance, in Earth's tropics \cite{Sobel2001,Seidel2020}. Thus, the area integrals become trivial. Conservation of mass in statistical equilibrium requires that $M_u A_u = -M_d A_d$. In the case of dry convection, $A_u = A_d$ \cite{Bjerknes1938,Singh2022a} implying that $|M| = M_u = -M_d$ and, if updrafts and downdrafts do equal amounts of pressure work, $\int_z B_u = -\int_z B_d$. In the third step, we invoke the equal area assumption of updrafts and downdrafts (this is an, admittedly, poor assumption for moist convection) and neglect CIN. Thus, we arrive at our desired approximation for the buoyancy flux (Equation \ref{eq:MB_simple}). CAPE is the convective available potential energy, i.e.~the part of the potential energy that is available to convert to kinetic energy. Finally, by substituting $Q_a = F_a A$ (Equation \ref{eq:Ebudget}) and Equation \ref{eq:MB_simple} into the simplified entropy budget (Equation \ref{eq:Wmax}), we arrive at
\begin{equation}
    \label{eq:HE}
     |M| \times \textrm{CAPE} = -\eta F_a \ .
\end{equation}
The above equation can be thought of as representing a convecting atmosphere as a heat engine in quasi-equilibrium. The engine is heated at the rate $F_s = -F_a$ (positive values for heating), but is not perfectly efficient and therefore does work at a rate $|M| \times \textrm{CAPE}$. Solving for the mass flux implied by the convective heat engine,
\begin{equation}
    \label{eq:M_HE}
    |M| = {-\eta F_a \over \textrm{CAPE}} \ .
\end{equation}

\subsubsection{The subcloud mass flux in radiative-convective equilibrium} \label{sec:subcloud}
Following \citeA{Emanuel1996}, we divide the system into a subcloud layer and a cloud layer. In the two-layer model, the cooling rate sums over both layers of the atmosphere:
\begin{equation}
    \label{eq:QA_2l}
    F_a = F_{a,sc} + F_{a,cl}
\end{equation}
The energy budget of the subcloud layer in radiative-convective equilibrium allows us to estimate the mass flux therein: 
\begin{align}
    A F_s + A F_{a,sc} - (A_u M_{u,sc} h_{sc} + A_d M_{d,cl} h_{cl}) &= 0 \nonumber \\
    \label{eq:H-budget}
    F_s + F_{a,sc} + |M_{sc}| (h_{cl} - h_{sc}) &= 0.
\end{align}
Here, the subscript ``sc'' indicates the subcloud layer, the subscript ``cl'' indicates the cloud layer, and $h = c_p T + g z + L q$ is the moist static energy (MSE). Therefore, $M_{u,sc}$ is the updraft mass flux from the sub-cloud layer and $M_{d,cl}$ is the downdraft mass flux from the cloud layer. To obtain Equation \ref{eq:H-budget}, we invoke mass conservation in statistical equilibrium ($A_u M_{u,sc} = -A_d M_{d,cl}$) and the equal area assumption of updrafts and downdrafts where $A_u + A_d = A$, implying that $|M_{sc}| = M_{u,sc} = -M_{d,cl}$ where $M_{sc}$ is the sub-cloud mass flux. Mixing equal parts of air between the two layers implies a net upward transport of MSE, though this is conditional upon MSE decreasing with height. Following \citeA{Emanuel1996}, we assume that the air parcels representative of the sub-cloud layer originate near the surface and those representative of the cloud layer originate near the tropospheric minimum in MSE, and are exchanged across the lifting condensation level (LCL). If the parcels conserve their MSE, the two-layer MSE difference is
\begin{align}
    \Delta h &= h_{sc} - h_{cl} \nonumber \\
    &= \Delta (c_p T) + \Delta (gz) + \Delta (Lq) \nonumber \\
    \label{eq:MSEdiff}
    &\approx c_p (T_s - T_{min}) + g (z_s - z_{min}) + L ( q_s - q_{min})
\end{align}
where the subscript ``$s$" indicates the near-surface, the subscript ``$min$" indicates the tropospheric minimum in MSE, $\Delta (c_p T) + \Delta (gz)$ is the dry static energy difference, and $\Delta (Lq)$ is the latent energy difference. Substituting $F_s = -F_a$ (Equation \ref{eq:Ebudget}) and Equation \ref{eq:QA_2l} into the sub-cloud energy budget (Equation \ref{eq:H-budget}), we solve for the mass flux from the sub-cloud layer:
\begin{equation}
    \label{eq:M_sc}
    |M_{sc}| = -{F_{a,cl}\over \Delta h} \approx -{F_a\over \Delta h}
\end{equation}
It is clear that the role of convection in radiative-convective equilibrium is to re-distribute latent and sensible heat. Here, we have assumed that cooling rates are small in the subcloud layer such that $F_{a,cl} \approx F_a$.
The potential for LTRH (i.e., $F_{a,sc}>0$) to violate QE could be explored in future work through Equations \ref{eq:Trad} and \ref{eq:M_sc}. The approximate form of $|M_{sc}|$ (Equation \ref{eq:M_sc}) follows from this simplification. $|M_{sc}|$ can be interpreted as the mass flux in radiative-convective equilibrium.  Mass continuity requires $|M_{sc}|\approx |M|$.

\subsection{Theory 2: convection in a zero-buoyancy world}\label{sec:ZBA}

As demonstrated by \citeA{S&O}, the vertical integral of cloud buoyancy taken relative to the clear-sky environment in convection-resolving model simulations is near zero. Based on this insight, they proposed a conceptual model of convection in which clouds are exactly neutrally-buoyant with respect to their environment. This assumption about typical cloud buoyancies is known as the \textit{zero-buoyancy approximation}. \citeA{Romps2014,Romps2016} introduced an analytical model of zero-buoyancy convection, in which the steady-state humidity and temperature fields of the atmosphere are determined by the turbulent interaction between ascending cloudy air and descending environmental air. The requirement that entraining clouds are neutrally buoyant with respect to the environment does not imply zero CAPE, which specifically depends on the buoyancy of a non-entraining parcel/plume relative to the mean environment. To see this, consider the formal definition of CAPE (Equation \ref{eq:cape(rho)}). Equations \ref{eq:cape(rho)} 
and \ref{eq:B} tell us that two variables are required to estimate CAPE: the mean density $\overline \rho$ and the non-entraining parcel/plume density $\rho$. It is straightforward to calculate $\rho$ given surface boundary conditions. What is needed to find CAPE, therefore, is a plausible environmental profile of $\overline \rho$.  The environmental $\overline \rho$ is equal to the density of an \emph{entraining} cloud in the zero-buoyancy approximation, and this is what allows a closed-form model of CAPE. The zero-buoyancy model of CAPE has been validated against convection-resolving model simulations \cite{Romps2016,Seeley2023}. 

Below, we re-derive the zero-buoyancy theory of CAPE \cite{Romps2016} in order to estimate its dependence on surface temperature and moisture. We begin by approximating the saturation specific humidity as 
\begin{equation}
    \label{eq:qstar}
    q^* = \frac{R_a}{R_v}\frac{e^*}{p},
\end{equation}
where $R_a$ is the specific gas constant of environmental air (everywhere assumed to be that of dry air) and $p$ is the total air pressure. Taking the vertical derivative of the natural log of $e^*$ (and using the definition of the lapse rate $\Gamma = -\partial_z T$), we obtain
\begin{eqnarray}
    &\partial_z e^* & = \partial_T e^* \partial_z T = -\frac{Le^*\Gamma}{R_vT^2}, \nonumber \\
    \label{eq:dzlogestar}
    &\partial_z \ln e^* &= -\frac{L\Gamma}{R_vT^2}.
\end{eqnarray}
The vertical derivative of the natural log of $p$ is obtained from hydrostatic balance and the ideal gas law:
\begin{equation}
\label{eq:dzlogp}
\partial_z\ln p = -\frac{g}{R_a T},
\end{equation}
where $g$ is gravity. Taking the vertical derivative of the natural log of $q^*$ and plugging in Equations \ref{eq:dzlogestar} and \ref{eq:dzlogp}, we obtain
\begin{eqnarray}
    \partial_z \ln q^* &=& \partial_z \ln e^* - \partial_z \ln p,\nonumber \\
     &=& \frac{g}{R_aT} - \frac{L\Gamma}{R_vT^2} = -\gamma. \label{eq:gamma}
\end{eqnarray}
where $\gamma$ is the moisture lapse rate (kg kg$^{-1}$ m$^{-1}$). The tropospheric moisture budget is obtained from the bulk-plume equations for convection in steady-state:
\begin{eqnarray}
    \label{eq:dzM}
    \partial_z M &=& e-d-c, \hspace{0.05 in} \textrm{where} \hspace{0.05 in} e=\varepsilon M \hspace{0.05 in} \textrm{and} \hspace{0.05 in} d = \delta M,\\
    \label{eq:dzMqstar}
    \partial_z (Mq^*)  &=& eq - dq^* - c, \hspace{0.05 in} \textrm{and} \\
    \label{eq:dzMq}
     -\partial_z(Mq)  &=& dq^* - eq + (1-\textrm{PE}) c.
\end{eqnarray}
$M$ is the convective mass flux (kg  m$^{-2}$ s$^{-1}$), $e$ and $d$ are the turbulent entrainment and detrainment rates (kg m$^{-3}$ s$^{-1}$) in which $\varepsilon$ and $\delta$ are fractional mixing efficiencies (m$^{-1}$), and $c$ is the condensation rate (kg m$^{-3}$ s$^{-1}$). $\textrm{PE}$ is the precipitation efficiency, defined as the fraction of condensates generated in updrafts at each height that are not re-evaporated in the environment. Per this definition, the gross evaporation is $(1-\textrm{PE}) c/M$ and the gross condensation minus gross evaporation is $\textrm{PE}c/M$. Re-evaporation is an irreversible source of entropy \cite{Emanuel2001} that is currently neglected in the heat engine model (Section \ref{sec:HE}). We make the following assumptions. The condensates not re-evaporated at each level ($\textrm{PE}c/M$) are immediately removed from the convective plume. The gross condensation represents a small fraction of the total updraft mass ($\partial_z M >> c)$. Invoking the latter assumption in Equation \ref{eq:dzM} gives
\begin{eqnarray}
    \label{eq:dzM_approx}
    \partial_z M &=& e-d \nonumber \\
    &=& (\varepsilon-\delta)M.
\end{eqnarray}
Expanding Equation \ref{eq:dzMqstar} with the chain-rule and solving for $\partial_zq^*$ (using Equation \ref{eq:dzM_approx}), we obtain
\begin{equation}
    \partial_z q^* = \varepsilon (q-q^*) - \frac{c}{M}.
\end{equation}
Doing the same to Equation \ref{eq:dzMq} to find $\partial_z q$,
\begin{equation}
    -\partial_z q = \delta (q^*-q) + (1-\textrm{PE}) \frac{c}{M}.
\end{equation}
The relative humidity is approximated as $\textrm{RH} = q/q^*$. Rearranging for $q$,
\begin{equation}
    \label{eq:qRq}
    q = \textrm{RH} q^*,
\end{equation}
and taking the vertical derivative of both sides, we obtain
\begin{equation}
    \label{eq:dzq-inter}
    \partial_z q  = q^* \partial_z \textrm{RH} + \textrm{RH}\partial_z q^*.
\end{equation}
We assume that vertical variations in RH are much smaller than those in specific humidity ($\partial_z \textrm{RH} << \partial_z q^*$), as is generally the case in Earth's troposphere. Invoking this assumption in Equation \ref{eq:dzq-inter} gives
\begin{equation}
\label{eq:dzq_inter2}
\partial_z q = \textrm{RH} \partial_z q^*.
\end{equation}
Using Equations \ref{eq:gamma}, \ref{eq:qRq}, and \ref{eq:dzq_inter2} to re-write Equations \ref{eq:dzMqstar} and \ref{eq:dzMq}, we obtain
\begin{eqnarray}
    \label{eq:6star}
    -\gamma q^* &=& \varepsilon (\textrm{RH}-1) q^* - \frac{c}{M}  \hspace{0.05in} \textrm{and}\\
    \label{eq:7star}
    \textrm{RH}\gamma q^* &=& \delta(1-\textrm{RH})q^* + (1-\textrm{PE}) \frac{c}{M}.
\end{eqnarray}
To solve for RH, we substitute $\frac{c}{M}$ from Equation \ref{eq:6star} into Equation \ref{eq:7star}. 
\begin{equation}
\label{eq:RH}
    \textrm{RH} = \frac{\delta + (1-\textrm{PE}) \gamma - (1-\textrm{PE}) \varepsilon}{\delta + \gamma - (1-\textrm{PE}) \varepsilon}
\end{equation}
We invoke the zero-buoyancy assumption \cite<>[]{S&O} to define the MSE ($h$) of the environment and the plume.
\begin{eqnarray}
    \label{eq:h}
    h &=& c_pT+gz+Lq \\
    \label{eq:hstar}
    h^* &=& c_pT+gz+Lq^*
\end{eqnarray}
In the zero-buoyancy assumption, ascending plumes are neutrally buoyant with respect to their environment. Strictly speaking, this means that their \textit{virtual} temperatures are the same. When virtual effects are neglected, as is done here, the plume and the environment possess the same temperature $T$ such that their moist static energies differ only by the differences in their specific humidities. $c_p$ is the specific heat of the atmosphere (assumed to be dry air, $c_{pa}$). Next, taking the vertical derivative of $h^*$ (and using $\Gamma = - \partial_z T$, $\partial_z q^* = -\gamma q^*$, and Equation \ref{eq:gamma}),
\begin{eqnarray}
    \partial_z h^* &=& -c_p \Gamma + g - L\gamma q^* \nonumber  \\
    \label{eq:dzhstar1}
    &=& g\Big(1 + \frac{Lq^*}{R_a T}\Big)-\Gamma\Big(c_p + \frac{L^2q^*}{R_vT^2}\Big).
\end{eqnarray}
It follows from Equation \ref{eq:dzM} that the vertical change in MSE flux with height for an entraining plume is
\begin{equation}
    \label{eq:Mhstar}
    \partial_z (Mh^*) = (\varepsilon h-\delta h^*)M.
\end{equation}
Using the chain rule to solve for $\partial_z h^*$ (and substituting Equations \ref{eq:dzM_approx}, \ref{eq:h}, and \ref{eq:hstar}):
\begin{eqnarray}
    \partial_z h^* &=& \varepsilon (h-h^*) \nonumber \\
    &=& \varepsilon (q-q^*)L \nonumber \\
    \label{eq:dzhstarb}
    &=& \varepsilon (\textrm{RH}-1)Lq^*.
\end{eqnarray}
To connect with the heat engine model, we assume that $M$ is constant with height. By Equation \ref{eq:dzM_approx}, this implies that $\varepsilon = \delta$. Next, we assume that RH and PE are also invariant. A consequence of this assumption is that, by Equation \ref{eq:RH}, the ratio of $\varepsilon$ and $\gamma$ is a constant. Following \citeA{Romps2016}, we combine the preceding constants into a ``bulk-plume parameter" that is, by design, invariant with height: 
\begin{equation}
\label{eq:a}
    a = \textrm{PE}\frac{\varepsilon}{\gamma}.
\end{equation}
The next step is to express the system of equations in the zero-buoyancy model in terms of this bulk-plume parameter. Relative humidity (Equation \ref{eq:RH}) simplifies to
\begin{equation}
    \label{eq:RHa}
    \textrm{RH} = \frac{1+a-\textrm{PE}}{1+a}.
\end{equation}
A trivial re-arrangement of Equation \ref{eq:a} yields $\varepsilon$:
\begin{equation}
    \label{eq:vareps}
    \varepsilon = \frac{\gamma a}{\textrm{PE}} 
\end{equation}
Substituting $\varepsilon$ (Equation \ref{eq:vareps}) and RH (Equation \ref{eq:RHa}) into Equation \ref{eq:dzhstarb},
\begin{equation}
\label{eq:dzhstar2}
    \partial_z h^* = -\frac{a}{1+a}\gamma Lq^*.
\end{equation}
Equating the expressions for the vertical gradient in saturation moist static energy (Equations \ref{eq:dzhstar1} and \ref{eq:dzhstar2}) and solving for $\Gamma$,
\begin{equation}
\label{eq:Gamma}
\Gamma = \frac{g}{c_p} \Big[ \frac{1+a+q^* L/(R_aT)}{1+a+q^* L^2/(c_pR_vT^2)} \Big].
\end{equation}
Equation \ref{eq:Gamma} is the temperature lapse rate set by entraining convection and is accurate when the water vapor mixing ratio is less than one. By neglecting virtual effects, the convective available potential energy (Equation \ref{eq:cape(rho)}) in steady-state becomes
\begin{equation}
    \label{eq:cape}
    \textrm{CAPE} \approx \int_{\text{LFC}}^{\text{LNB}} g\frac{T-\overline T}{\overline T} dz,
\end{equation}
where $T$ is the temperature of a pseudo-adiabatic parcel and $\overline T$ is the mean environmental temperature. We obtain $T$ and $\overline T$ by integrating the zero-buoyancy model vertically with $a=0$ and $a\geq0$, respectively. PE and $a$ are prescribed constants. Given the precipitation efficiency and mean relative humidity from simulations, an appropriate input value of $a$ can be diagnosed via Equation \ref{eq:RHa}. $a$ controls the moist coupling between convective plumes and environmental air \cite<for further discussion, see>[]{Seeley2023}. For non-entraining convection ($a=0$), $\Gamma$ equals the moist adiabatic lapse rate, $\Gamma_m$. As $a$ increases, the tighter coupling between the entraining plume and the environment forces $\Gamma$ apart from $\Gamma_m$ \cite{Seeley2023}, permitting more CAPE in steady-state.

The zero-buoyancy model shows a peak in CAPE at intermediate surface temperatures (Figure \ref{fig:breakdown}a) because the temperature difference $\Delta T = T-\overline T$ (Equation \ref{eq:cape}) between adiabatic parcels and the mean environment becomes negligible when the saturation specific humidity approaches the extremes of zero or one. Since CAPE is directly proportional to $\Delta T$, CAPE is near zero in both the dry and moist limits, with a maximum in between \cite{Seeley2023}. From a more technical perspective, \citeA{S&O} explain that $\Delta T$ is inversely proportional to a pseudo-heat capacity $\beta = c_p + L^2q^*/(R_v T^2)$. Since the heat capacity of dry air is held constant, $1/\beta$ primarily decreases with increasing moisture. Combining these insights, \citeA{Romps2016} demonstrate that CAPE starts to decrease with rising surface temperatures when $\beta > 2 c_p$ throughout the troposphere (see their Figure 9), meaning further increases in moisture reduce $\Delta T$. 

To summarize, the system of equations for a convecting atmosphere in radiative-convective equilibrium under the zero-buoyancy approximation are $q$ and $q^*$, $\gamma$, RH, $\Gamma$, and CAPE. The final forms of the equations assume that the convective mass flux, the relative humidity of the environment, and precipitation efficiency are constant with height. The thermodynamic constants and their units and values in ``Earth-like" and ``Titan-like" conditions are given in \ref{app:A} and are also assumed to be constant with height.

\subsection{The equilibrium condition}\label{sec:condition}
\begin{figure*}
\noindent\includegraphics[width=0.95\textwidth,angle=0]{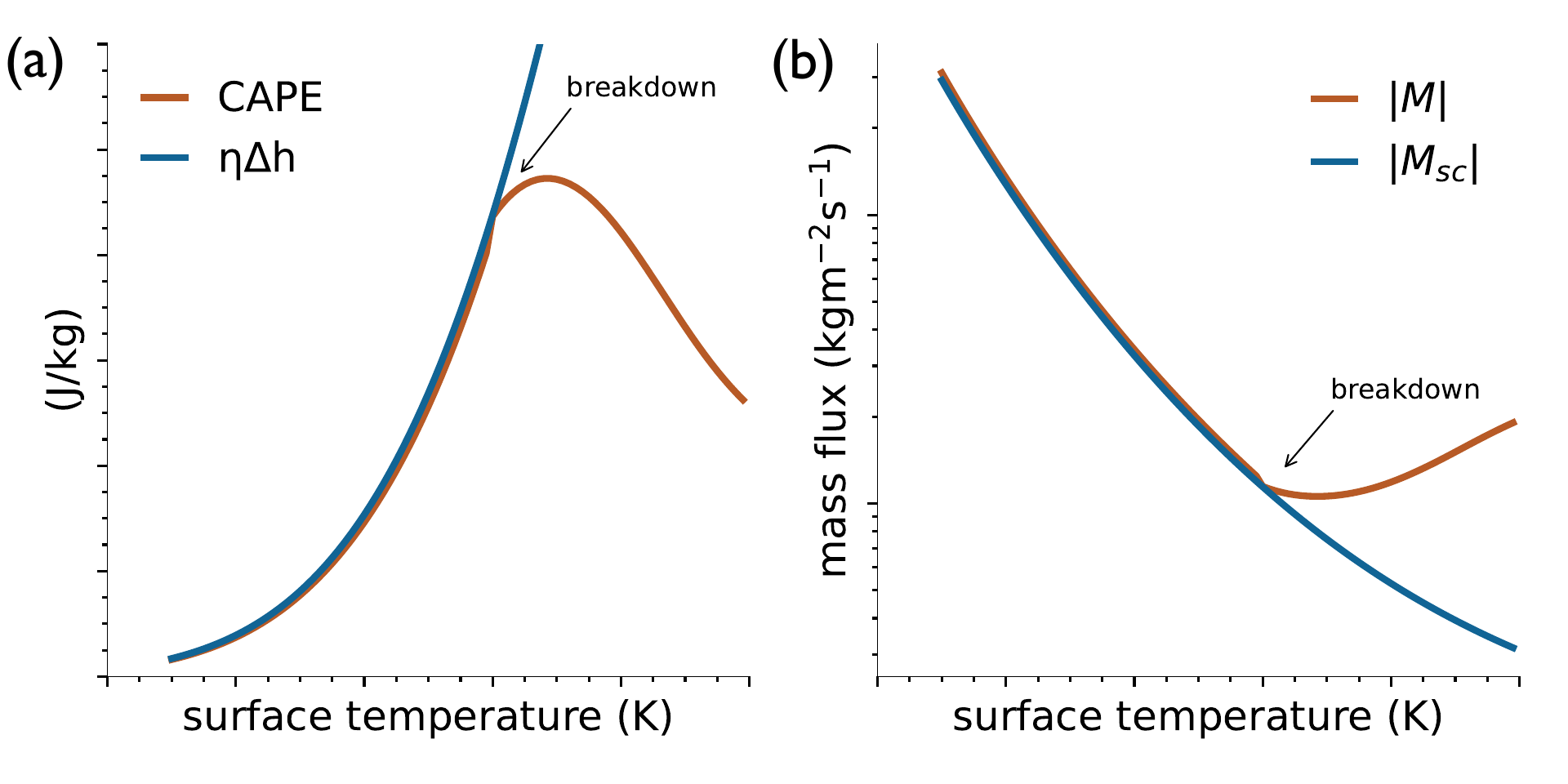}
  \caption{Schematic illustration of the breakdown of convective quasi-equilibrium with increasing temperature. (a) Comparison of convective available potential energy (CAPE; J kg$^{-1}$) and the vertical MSE difference ($\Delta h$; J kg$^{-1}$) times the heat engine efficiency $\eta$ as a function of surface temperature. (b) Comparison of the heat engine mass flux $|M| \propto 1/\textrm{CAPE}$ and the subcloud mass flux $|M_{sc}|\propto 1/\Delta  h$ as a function of surface temperature. Quasi-equilibrium breaks down when $\eta \Delta  h > \textrm{CAPE}$, as this indicates $|M_{sc}|<|M|$.} \label{fig:breakdown}
  \centering
\end{figure*}

\citeA{Emanuel1996} equate $|M|$ (Equation \ref{eq:M_HE}) and $|M_{sc}|$ (Equation \ref{eq:M_sc}) to derive an expression for CAPE. This is clearly necessary for statistical equilibrium, in which there is no net vertical transport of mass. Doing so, we find that QE convection requires a conversion of the vertical MSE difference ($\Delta h$) into CAPE at an efficiency $\eta$:  
\begin{equation}
    \label{eq:M=Msc}
    |M| \approx |M_{sc}| \rightarrow \eta \Delta  h \approx \textrm{CAPE}.
\end{equation}
Henceforward, Equation \ref{eq:M=Msc} is called the \textit{equilibrium condition}. To estimate CAPE and $\Delta  h$, we instead use a zero-buoyancy model of convection (see Section \ref{sec:ZBA}). It's important to note that the CAPE predicted by the zero-buoyancy model represents a steady-state storage of buoyancy, not the rate of CAPE generation and destruction by radiation and convection. 

We pose the following question: Is there a regime where $|M|$ and $|M_{sc}|$ don't equal one another, and if so, what does that imply about convection? In such a regime, QE convection would be incompatible with radiative-convective equilibrium. We argued previously that $|M|\propto 1/\textrm{CAPE}$ (Equation \ref{eq:M_HE}) and $|M_{sc}|\propto1/\Delta  h$ (Equation \ref{eq:M_sc}), so any constraint on mass fluxes naturally applies to CAPE and $\Delta h$. This is depicted schematically in Figure \ref{fig:breakdown}. Of course, $\Delta  h$ can increase without restriction (up to the limit of a steam atmosphere) as moisture increases by the Clausius-Clapeyron relation (Figure \ref{fig:breakdown}a). This implies that $|M_{sc}|$ decreases monotonically (Figure \ref{fig:breakdown}b). However, convection in the zero-buoyancy model has a peak in CAPE at intermediate surface temperatures \cite<Figure \ref{fig:breakdown}a;>[]{Seeley2023} due to the increasing influence of latent heat on the ``effective" heat capacity of the troposphere \cite{Romps2016}, implying a lower bound on $|M|$ but no upper bound (Figure \ref{fig:breakdown}b).  If $|M|<|M_{sc}|$, the heat engine of the cloud layer would be over-driven, this would likely reduce  CAPE, and the system would adjust to QE.  This adjustment process implies $|M|\approx |M_{sc}|$ is a characteristic of QE, which is the condition invoked by others to constrain CAPE \cite{Renno1996,Emanuel1996}.  Clearly, our theory also predicts $|M| > |M_{sc}|$ at surface temperatures above the threshold where $\eta \Delta h >$ CAPE (Figure \ref{fig:breakdown}b).  In this case, the subcloud layer cannot meet the mass flux demand required of the cloud layer heat engine, the layers decouple, and the cloud is cut off from the moisture source --  steady, QE convection cannot exist in this regime. Lacking convection, CAPE would build over time until $|M| \approx |M_{sc}|$, triggering convection. Our ``QE-breakdown" hypothesis for the emergence of RO convection is simply:
\begin{equation}
\eta \Delta h > \text{CAPE} .  
\end{equation}

\section{Testing the theory against convection-resolving model simulations}\label{sec:SW21}
\begin{figure*}
\noindent\includegraphics[width=0.95\textwidth,angle=0]{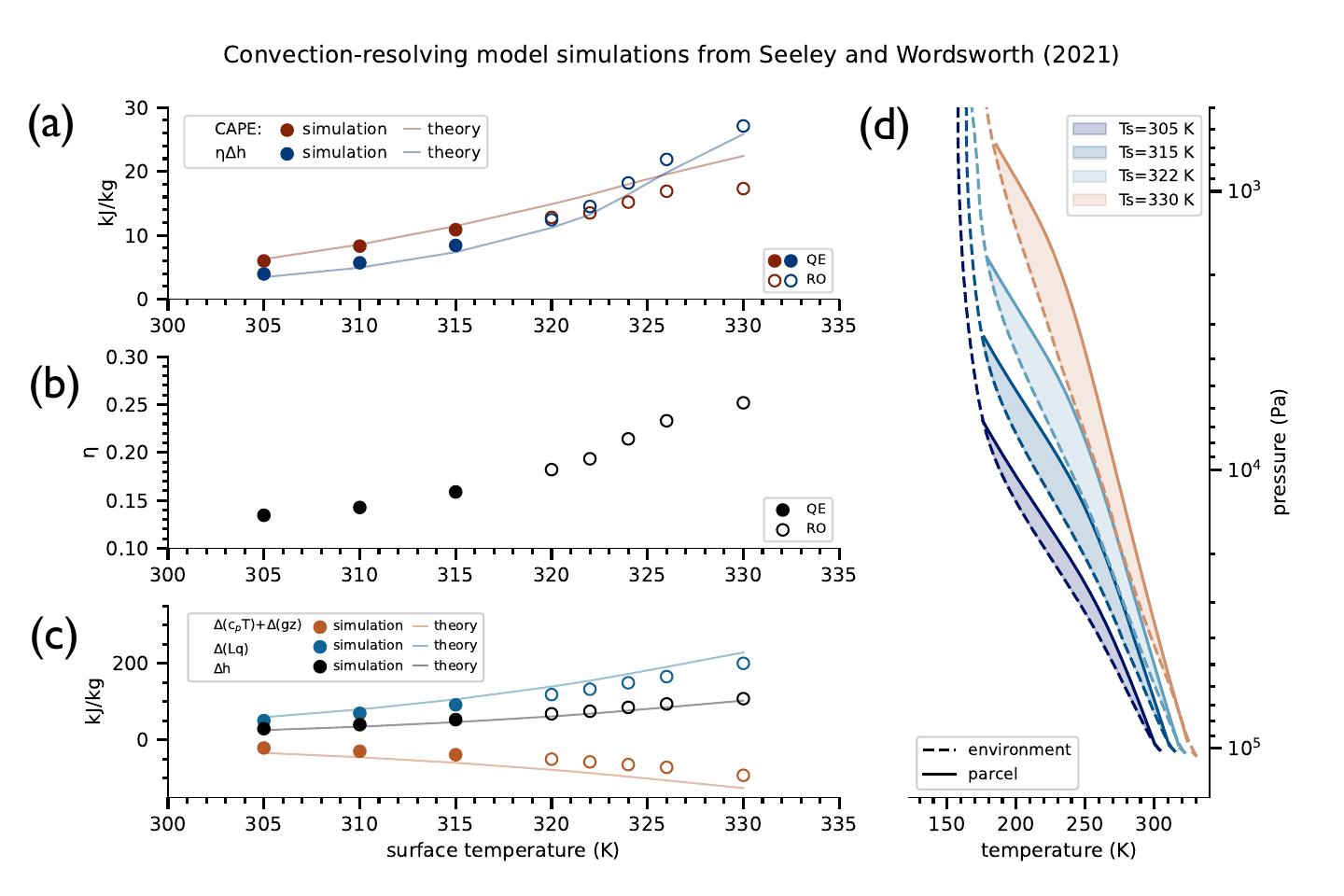}
  \caption{(a) Comparison of convective available potential energy (CAPE; J kg$^{-1}$) and the vertical MSE difference ($\Delta h$; J kg$^{-1}$) times the heat engine efficiency $\eta$ as a function of surface temperature. (b) Comparison of $\eta$ (blue) with the maximum theoretical efficiency, $\eta_{max}$, as a function of surface temperature. (c) Dry $\Delta(c_p T)$ and moist $\Delta (Lq)$ contribution to the MSE difference of upward and downward plumes across the LCL as a function of surface temperature. (d) Temperature difference of the environment and an adiabatically-lifted surface parcel in the simulation, given as a proxy for the steady-state CAPE. In (a)-(c), circular markers are diagnosed values from the fixed sea surface temperature simulations of \citeA{Seeley2021}.} \label{fig:sw21}
  \centering
\end{figure*}

We now test the zero-buoyancy heat engine theory of convection against fixed surface temperature simulations from \citeA{Seeley2021} performed with the convection-resolving model DAM \cite{Romps2008}. We use separate procedures to obtain estimates for $\eta$, $\Delta  h$, and CAPE from the theory and from the simulations, which are described below.  Some elements of the theory must be diagnosed from the simulations, so by ``testing" we mean that we are establishing consistency between QE breakdown of the heat engine model and the onset of the RO state in simulations.

To calculate the steady-state CAPE in the simulations as a function of surface temperature (Figure \ref{fig:sw21}a), we use the standard formula that includes virtual effects (Equation \ref{eq:cape(rho)}). The temperature of the ``adiabatic" parcel varies with height in accordance with the conservation of the sum of MSE and CAPE \cite<i.e., MSE+CAPE;>[]{Romps2015,Marquet2016}. We parameterize condensed water loss from the parcel as exponential decay over a length scale of 5 km following \citeA{Seeley2023}. The implementation of the MSE+CAPE parcel lifting method is detailed in \ref{app:C} following \citeA{Romps2015}. 

To obtain a theoretical prediction for CAPE, we initialize the zero-buoyancy model with the temperature and pressure of a near-surface parcel at the LCL. The precipitation efficiency (PE) and the relative humidity (RH) as a mass-weighted tropospheric mean are determined from the simulation. We diagnose the precipitation efficiency as the ratio of the surface precipitation rate ($P_s$; kg m$^{-2}$ s$^{-1}$) to the vertically-integrated sink of water vapor associated with phase changes (SI; kg m$^{-2}$ s$^{-1}$) following \citeA{Sui2007}: 
\begin{equation}
\label{eq:PE-diag}
\text{PE} = \frac{P_s}{\text{SI}}.
\end{equation}
By substituting PE and RH into Equation \ref{eq:RHa}, we obtain a self-consistent estimate for the bulk-plume parameter in the simulation: 
\begin{equation}
    a = \max\Bigg(\frac{1-\text{RH}-\text{PE}}{\text{RH}-1},0\Bigg).
\end{equation}
$a$ quantifies the strength of the coupling between entraining plumes and their environment, taking into account both entrainment and re-evaporation. The constraint that $a\geq0$ follows from the requirement that $\text{RH}\geq 1-\text{PE}$ \cite{Romps2014}. When $a=0$, $\text{RH}=1-\text{PE}$ and environmental temperatures follow the moist adiabat. For those interested, Figure \ref{fig:conv-params}b-d in \ref{app:B} shows PE, RH, and $a$ from the convection-resolving model of \citeA{Seeley2021}. Representative parameter values of the quasi-equilibrium state in the convection-resolving model ($\text{RH}=0.82$, $\text{PE}=0.27$, and $a=0.5$) are used to evaluate the zero-buoyancy model as a function of temperature. Figure \ref{fig:sw21}a shows the theoretical and simulated CAPE. The simulated CAPE steadily increases in the convection-resolving model experiments of \citeA{Seeley2021} from 6 kJ/kg to 17 kJ/kg over the surface temperature range of 305-330 K. The growth in CAPE is explained by Figure \ref{fig:sw21}d, which compares the environmental temperature to that of a surface parcel displaced adiabatically to the level of neutral buoyancy. CAPE is proportional to shaded area between the two curves in Figure \ref{fig:sw21}d that represents the temperature difference between the environment and the parcel, where it can be seen that the size of the shaded area increases with surface warming in the simulations.

The heat engine efficiency $\eta$ (Equation \ref{eq:eta}) depends on the mean temperature at which frictional dissipation occurs ($T_d$) and the mean inverse temperatures at which radiation is absorbed at the surface ($1/T_s$) and emitted from the atmosphere ($1/T_a$). The maximum efficiency of the convective heat engine ($\eta_{max}$; Figure \ref{fig:sw21}b) would be achieved if frictional dissipation occurs only at the surface and the net emission level corresponds to the tropopause, $\eta_{max} = (T_s - T_{trp})/T_{trp}$ , where $T_{trp}$ is an assumed tropopause temperature of 200 K.  $T_s$ and $T_{trp}$ clearly represent the warmest and coldest points of the system, respectively. Over the surveyed surface temperature range, $\eta_{max}$ takes values between 52-65\%. Rather than assuming the maximum efficiency, we make the reasonable assumption that most of the frictional dissipation occurs between the surface and the effective emission level such that $T_d = (T_s + T_a)/2$.  Lacking a theory for the radiative cooling of the atmosphere ($Q_a$) to diagnose $T_a$ (Equation \ref{eq:Trad}), we instead obtain it from the model output of \citeA{Seeley2021}, which yields our estimate for $\eta$ with realistic radiation (Figure \ref{fig:sw21}b). We find that $T_a$ ranges from 258 K to 272 K (Figure \ref{fig:conv-params}a) in the simulations.  Since the heat engine efficiency is sensitive to the radiation, we treat the simulation values as the ``theoretical" prediction for $\eta$. The predicted efficiency is almost 15\% at 305 K, and increases with surface warming up to 25\% at 330 K. Raising the surface temperature drives more water vapor into the atmosphere, which increases radiative absorption at infrared and visible wavelengths. The spectral region over which the present-day atmosphere is transparent to infrared radiation (i.e., the water vapor window) begins to close at surface temperatures greater than 300 K, and becomes fully opaque at 320 K due to water vapor continuum absorption \cite{Koll2018}. Thus, the radiative properties of water vapor reduce $T_a$ with surface warming above 320 K (Figure \ref{fig:conv-params}a), which along with increasing surface temperatures further increases $\eta$. 

The vertical difference in MSE, $\Delta h$ (Equation \ref{eq:MSEdiff}), dry static energy, $\Delta (c_p T) + \Delta (gz)$, and latent energy, $\Delta (Lq)$ are displayed in Figure \ref{fig:sw21}c. Our method of evaluating $\Delta h$ in the simulations and the theory is described in Section \ref{sec:subcloud}. There is no boundary layer in the zero-buoyancy model (Section \ref{sec:ZBA}), so we take the MSE of the sub-cloud layer to be that of the lowest atmospheric layer in the simulation. Then, we take the MSE of the cloud layer to be the MSE minimum in the zero-buoyancy model. The theoretical $\Delta h$ slightly underestimates the simulated values, which increase from 29 kJ/kg to 107 kJ/kg over the experimental range of surface temperatures (Figure \ref{fig:sw21}c). The steady growth in $\Delta h$ reflects a competition between the increasing latent energy difference and the decreasing dry static energy difference. The vertical latent energy difference represented by $\Delta(Lq)$ is dependent on temperature through the Clausius-Clapeyron relation, explaining its positive rise from 50 kJ/kg to 200 kJ/kg (Figure \ref{fig:sw21}c). A larger and more negative vertical difference in dry static energy of -21 kJ/kg to -92 kJ/kg develops because of the expansion of the troposphere with surface warming. This expansion yields a larger geopotential energy difference in which $\Delta (g z)<0$, and this cancels out the positive growth in $\Delta (c_p T)>0$ (not shown). Overall, the growth in $\Delta h$ is fueled by latent component (Figure \ref{fig:sw21}c).

The equilibrium condition implies that there should be a breakdown of steady, QE convection if CAPE $< \eta \Delta  h$. In the simulations of \citeA{Seeley2021}, we see that CAPE$<\eta \Delta h$ above a surface temperature of 320 K (circles in Figure \ref{fig:sw21}a), which coincides with the transition into the RO regime.  The zero-buoyancy heat engine model (lines in Figure \ref{fig:sw21}a) predicts the breakdown of QE convection within 5 K of the onset of RO convection, using QE values of the convective parameters from the simulation (Figure \ref{fig:conv-params}b-d). Their simulations show an increase in $\eta \Delta  h$ with surface warming largely due to the increase in $\Delta(Lq)$. Convection transports more energy per unit mass of cloudy air in warmer climates. The efficiency of the convective heat engine also increases in warmer climates partly due to the increase in surface temperature and partly due to the radiative properties of water vapor, which shift the emission level upward. In summary, the convection-resolving simulations seem to support our zero-buoyancy heat engine hypothesis: RO convection emerges due to a breakdown in QE convection, which is caused by radiative and thermodynamic effects of increases in water vapor.

\section{RO states exist in a single-column model of radiative-convective equilibrium}

We've demonstrated an important contradiction between quasi-equilibrium convection and radiative-convective equilibrium at high surface temperatures. This led us to a novel explanation for RO emergence. The heat engine (Section \ref{sec:HE}) and zero-buoyancy (Section \ref{sec:ZBA}) theories of convection posit only the existence of an ensemble of convective plumes that are in steady-state. The heat engine theory acknowledges the potential existence of spatial inhomogeneities in, for instance, radiation, but either averages over them or makes simplifying assumptions to arrive at the useful bulk quantities. In the zero-buoyancy theory, the properties of the environment (this being the sub-saturated downdrafts) including temperature and relative humidity are determined by their mutual interaction with the cloudy updrafts. While this conceptual model is based on three-dimensional reality (Figure \ref{fig:HE}), the assumption of horizontal homogeneity in temperature of the zero-buoyancy model allows for the governing equations to be evaluated as a function of height alone. Indeed, the only consideration of horizontal variations is implicit in the humidity difference between ascending and descending plumes. These plumes are not spatially resolved but instead their bulk properties are diagnosed from the large-scale environmental variables. Hydrostatic climate models parameterize convective processes in a single vertical dimension using this bulk-plume approach. 

We see no a-priori reason why a single-column model of radiative-convective equilibrium should not exhibit RO dynamics at high surface temperatures so long as the convection scheme represents steady-state ascending and descending motion by such a bulk-plume parameterization. As in the case with resolved convection, we should expect RO dynamics in a single-column model of radiative-convective equilibrium if $\text{CAPE}< \eta \Delta h$. To test these ideas, we reproduce the basic experimental setup of \citeA{Seeley2021} in a single-column climate model with a bulk-plume parameterization of convection.

\subsection{Model and Methods}\label{sec:methods}

We use a version of the ECHAM6 general circulation model \cite{Stevens2013} in single-column mode that has been modified to allow water vapor to comprise a significant fraction of the atmospheric mass \cite{Popp2015}. The single-column model is forced only by surface heating and radiative cooling (i.e., radiative-convective equilibrium) and has separate schemes for radiation, convection, clouds, and turbulent fluxes.

\subsubsection{Base experiment}\label{subsec:base}

In our base experiment, we run simulations over a temperature range from 290 K to 360 K. The insolation is set 10\% higher than the present-day value and is temporally fixed (no diurnal or seasonal cycle). We set the column latitude to 38$^{\circ}$N, where the globally-averaged insolation is the same as the local value. Clouds are the only source of time-varying planetary albedo. The atmosphere is composed \textit{only} of nitrogen, oxygen, carbon dioxide, and water. The molar mixing ratio of carbon dioxide is set to 354 ppmv and is uniform with height. We use a time step of 60 seconds and run the simulations for approximately 200 years. The surface temperature of a 1 m mixed-layer ocean with an albedo of 0.07 is fixed at every time step through the use of an artificial surface heat sink. 

\subsubsection{Thermodynamics}
The version of ECHAM6 that we use accounts for the contribution of water vapor to the total pressure, density, and heat capacity of the atmosphere. The model uses an empirical formula for the saturation vapor pressure of water over liquid and ice, respectively:
\begin{align}
e^* (\text{Pa}) & = \exp\Big(c_1/T + c_2 + 10^{-2} c_3 T + 10^{-5} c_4 T^2 + c_5 \ln(T)\Big) \\
\text{where} \: c_1,c_2,c_3,c_4,c_5&=
    \begin{cases}
        -6024.5282,29.32707,1.0613868,-1.3198825,-0.49382577 & \text{if } T \leq 273.15 \: \text{K} \\
        -6096.9385,21.2409642,-2.711193,1.673952,2.433502 & \text{if } T > 273.15 \: \text{K}
    \end{cases}\nonumber
\end{align}
The constants take different values depending on whether the temperature (in units of Kelvin) is above or below the triple point temperature.

\subsubsection{Radiation}
Shortwave and longwave radiation is resolved into 14 and 16 spectral bands by the Rapid Radiative Transfer Model for General Circulation Models \cite<RRTMG;>[]{Iacono2008}. RRTMG uses the correlated-k method under the two-stream approximation, where each band is further sub-divided on the basis of the strength of molecular absorption features. This results in a spectrally-integrated radiative heating rate with 140 pseudo-wavelengths in the longwave and 112 pseudo-wavelengths in the shortwave \cite{Giorgetta2013}. By default, the radiation calculation is performed once hourly. All forms of water besides precipitation are accounted for in the radiation calculation. As in \citeA{Popp2015}, we use an exponential extrapolation of all molecular absorption coefficients in the longwave and the water self-broadened absorption coefficients in the shortwave for temperatures above which no data in the original model exists. The effect of pressure broadening by water vapor is neglected.

\subsubsection{Convection}\label{subsec:convection-echam6}
Convection is represented by the \citeA{Nordeng1994} bulk-plume scheme, which parameterizes turbulent entrainment and detrainment of air between updrafts, downdrafts, and the environment. The scheme distributes energy, moisture, and momentum through the column under the assumption that the cumulus ensemble is in steady-state. Downdrafts are initialized at the level of free sinking with the properties of a mixture of cloudy and saturated environmental air at their wet bulb temperature \cite{Giorgetta2013}. The level of free sinking is defined as the highest location where said mixture is negatively buoyant with respect to the environment. The downdrafts remain saturated by re-evaporating condensates produced by the convective updrafts. The initial downdraft mass flux is assumed to be directly proportional to the initial updraft mass flux \cite{Nordeng1994}. Note that the zero-buoyancy model of \citeA{Romps2016} neglects condensate loading of updrafts (liquid water is instantly removed) and the effect of downdrafts associated with re-evaporation; it does, however, account for adiabatic subsidence (Section \ref{sec:ZBA} ). For saturated updrafts undergoing pseudo-adiabatic ascent, the bulk-plume equations of \citeA{Nordeng1994} are the same as in the zero-buoyancy model of convection (Equations \ref{eq:dzMqstar} and \ref{eq:dzM_approx}). The convection scheme has a QE-type closure \cite{Neelin2000} so that the cloud-base mass flux $M_{cb} \propto \textrm{CAPE}/\tau$ where $\tau = 2$ hours is an assumed adjustment time for CAPE \cite{Giorgetta2013}. To ensure numerical stability, the permitted range of cloud base mass fluxes is $10^{-10}$--$1$ kg m$^{-2}$s$^{-1}$ and convective temperature tendencies are limited to 0.05 K/s. We set the entrainment rate in the convection scheme to 0.1 km$^{-1}$, following \citeA{Popp2015} and \citeA{fsa23a}. The updraft mass flux is assumed to be constant up to a critical height set by the buoyancy of entraining convection, above which updrafts are only permitted to detrain \cite{Mobis2012}. The downdraft mass flux is assumed to be constant with height \cite{Nordeng1994,Tiedtke}. The trigger for shallow convection is based on the buoyancy of adiabatically lifted parcels relative to the environment at the cloud base \cite{Mobis2012}.

The definition of QE varies between contexts, as is elegantly described in a review article by \citeA{Yano2012}. As stated above, the single-column model utilizes a QE-type convection scheme. In the context of a convection scheme, ``QE” has a specific meaning. Convection schemes require a closure, which is an assumption that enables a prediction of the \textit{instantaneous} convective mass flux. These schemes conceptualize convection as a rapid relaxation process that destroys CAPE. Hence, QE is applied as a concept of balance to the CAPE budget \cite<i.e., the time rate of change of CAPE;>[]{Yano2012}, which describes the time rate of change of CAPE due to radiative and convective processes; the former generates CAPE, while the latter dissipates it. In doing so, the rates of CAPE generation and destruction are assumed to be equal. As discussed, the solution for the instantaneous mass flux in the QE convection scheme is proportional to $\textrm{CAPE}$ \cite{Giorgetta2013}. Notably, however, the \textit{steady-state} heat engine mass flux (Equation \ref{eq:M_HE}) is inversely proportional to CAPE. Albeit a technical detail, it is surprising that applying QE reasoning to the entropy budget and the CAPE budget results in formulations of the convective mass flux that seem at odds. 

\begin{figure*}
\noindent\includegraphics[width=0.95\textwidth,angle=0]{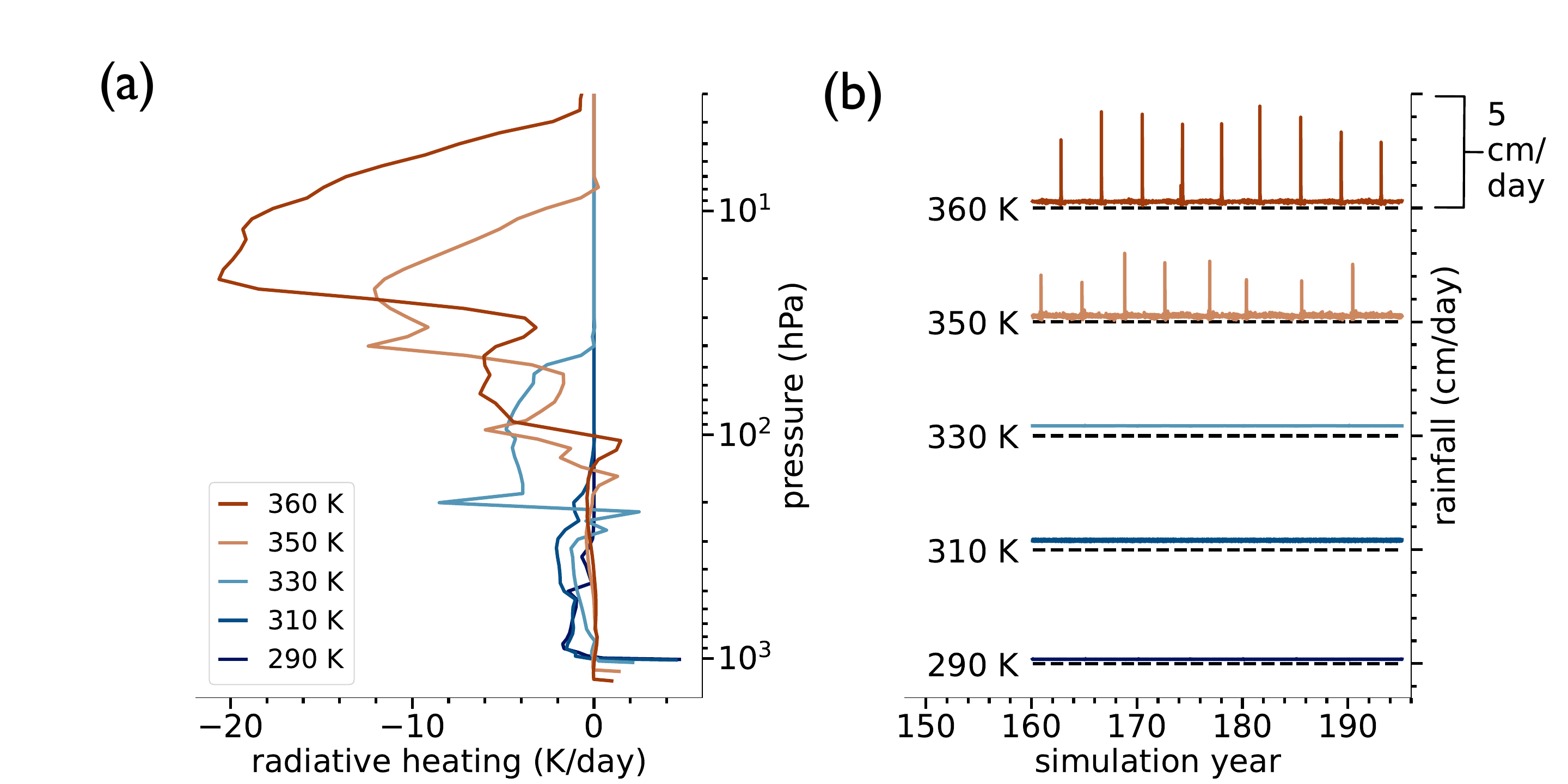}
\caption{From the base experiment in ECHAM6, (a) net radiative temperature tendency (K/day) and (b) time series of the surface precipitation rate (cm/day). Values in (a) are temporally-averaged over the multi-decadal period in (b). Time is given in years since the start of the simulation. Precipitation rates at each surface temperature are vertically offset by 5 cm/day and a zero rainfall contour is given for reference as a dashed black line.
\label{fig:rain-transition}}
\end{figure*}

\subsubsection{Clouds}

The large-scale cloud scheme has separate prognostic equations for the vapor, liquid, and ice phases of water, a modified microphysics scheme based on \citeA{Lohmann1996}, and diagnostic cloud cover \cite{Sundqvist}. Sources and sinks of water from non-local transport processes such as convection and turbulence and local processes such as condensation, evaporation, deposition, sublimation, precipitation formation, re-evaporation of rain and sublimation of snow are included in the prognostic equations \cite{Giorgetta2013}. The amount of rain re-evaporation at each height is proportional to the local saturation deficit - that is, the difference between the saturation and environmental water vapor mixing ratio \cite{Lohmann1996}. Cloud fraction is diagnosed at each time step as a function of the environmental relative humidity \cite{Stevens2013}. The cloud fraction is calculated only if the relative humidity is greater than a threshold value, which monotonically decreases with height. Condensational growth or evaporative decay of cloud droplets is conditional on whether the relative humidity is above or below this threshold \cite{Giorgetta2013}.

\subsubsection{Surface fluxes and eddy diffusion}
Sensible and latent heat fluxes at the surface are determined by the standard bulk-exchange formulas. Vertical turbulent mixing is parameterized using the eddy-diffusivity approach of \citeA{Brinkop1995}.

\subsubsection{Precipitation efficiency} \label{sec:PE-echam}
We diagnose the precipitation efficiency in ECHAM6 in accordance with Equation \ref{eq:PE-diag},
where
\begin{equation*}
    P_s = P^{ls}_s + P^{u}_s
\end{equation*}
and
\begin{equation*}
\text{SI} = \text{SI}^{ls} + \text{SI}^{u}.
\end{equation*}
The superscripts ``ls" and ``u" refer to the large-scale environment and convective updrafts, respectively. $P_s^{ls}$ and  $P_s^{u}$ are surface precipitation rates, which are determined separately by the large-scale cloud scheme and the convection scheme. Note that convection-resolving models make no distinction between $P_s^{ls}$ and $P_s^u$. $\text{SI}^{ls}$ and $\text{SI}^{u}$ are the vertically-integrated gross sinks of water vapor associated with phase changes in the large-scale environment and in convective updrafts, respectively. The vertically-integrated gross sink of water vapor in the large-scale environment from the surface (SFC) to the tropopause (TRP) is
\begin{equation*}
    \text{SI}^{ls} = \int_{\text{SFC}}^{\text{TRP}} \Big(\frac{\partial q}{\partial t}\Big)^{ls} \: \frac{dp}{g},
\end{equation*}
where the gross large-scale sink of water vapor is
\begin{equation*}
    \Big(\frac{\partial q}{\partial t}\Big)^{ls} = \dot q_{cnd} + \dot q_{dep}+ \dot q_{tbl}+ \dot q_{tbi} < 0. 
\end{equation*}
Here, $\dot q_{cnd}$ and $\dot q_{dep}$ are the condensation and deposition rates of water vapor, and $\dot q_{tbl}$ and $\dot q_{tbi}$ are the rates of cloud condensate generation (liquid and ice, respectively) through turbulent fluctuations \cite{Giorgetta2013}. The vertically-integrated sink of water vapor in convective updrafts from the cloud base (CB) to the cloud top (CT)
\begin{equation*}
    \text{SI}^{u} = -\int_{\text{CB}}^{\text{CT}} M_u \frac{\partial q_u}{\partial z} \: dz,
\end{equation*}
where $M_u$ is the updraft mass flux and $\frac{\partial q_u}{\partial z}<0$ is the gross vertical change in updraft specific humidity due to condensation and/or deposition. This definition of $\text{SI}^{u}$ excludes other processes that alter $q_u$ with height but that are not associated with phase changes, such as entrainment. Downdrafts do not generate condensates in ECHAM6; the parameterized effect of downdrafts is to evaporate the condensates produced in updrafts in order to maintain their saturated descent, thereby reducing the overall convective precipitation. 

\subsection{RO emergence is consistent with a breakdown of QE}

To date, RO states have only been simulated in Earth models with resolved convection \cite{Seeley2021,Dagan2023,Song2024}. Figure \ref{fig:rain-transition}b clearly shows that the simulated climate in our base experiment transitions into the RO state at surface temperatures around 350 K. At cooler temperatures, precipitation is steady with a mean value of 2-6 mm/day; these simulations are in the QE regime. RO states exhibit episodic precipitation with intensities up to 5 cm/day that repeat every $O(100-1000)$ days and that are relatively short-lived (10-30 days). The storm duration is calculated as the number of contiguous days where the precipitation rate is above the mean value. Here, we stress that the storm duration and frequency are inconsistent with previous estimates for hothouse climates. Convection-resolving models find that the storms last several hours and reappear in a matter of days \cite{Seeley2021,Dagan2023,Song2024}. We speculate that these orders-of-magnitude differences arise from the parameterizations in our one-dimensional model. We cannot rule out the possibility that the driving physics of the RO regime are different between one-dimensional and three-dimensional models. Our partial replication of the RO regime in a single-column model underscores the important limitations of simpler parameterized climate models. The accuracy of the single-column model might be improved by tuning or overhauling existing parameterizations (see \ref{app:B}), a task that we leave to future work. 

\begin{figure*}
\noindent\includegraphics[width=0.95\textwidth,angle=0]{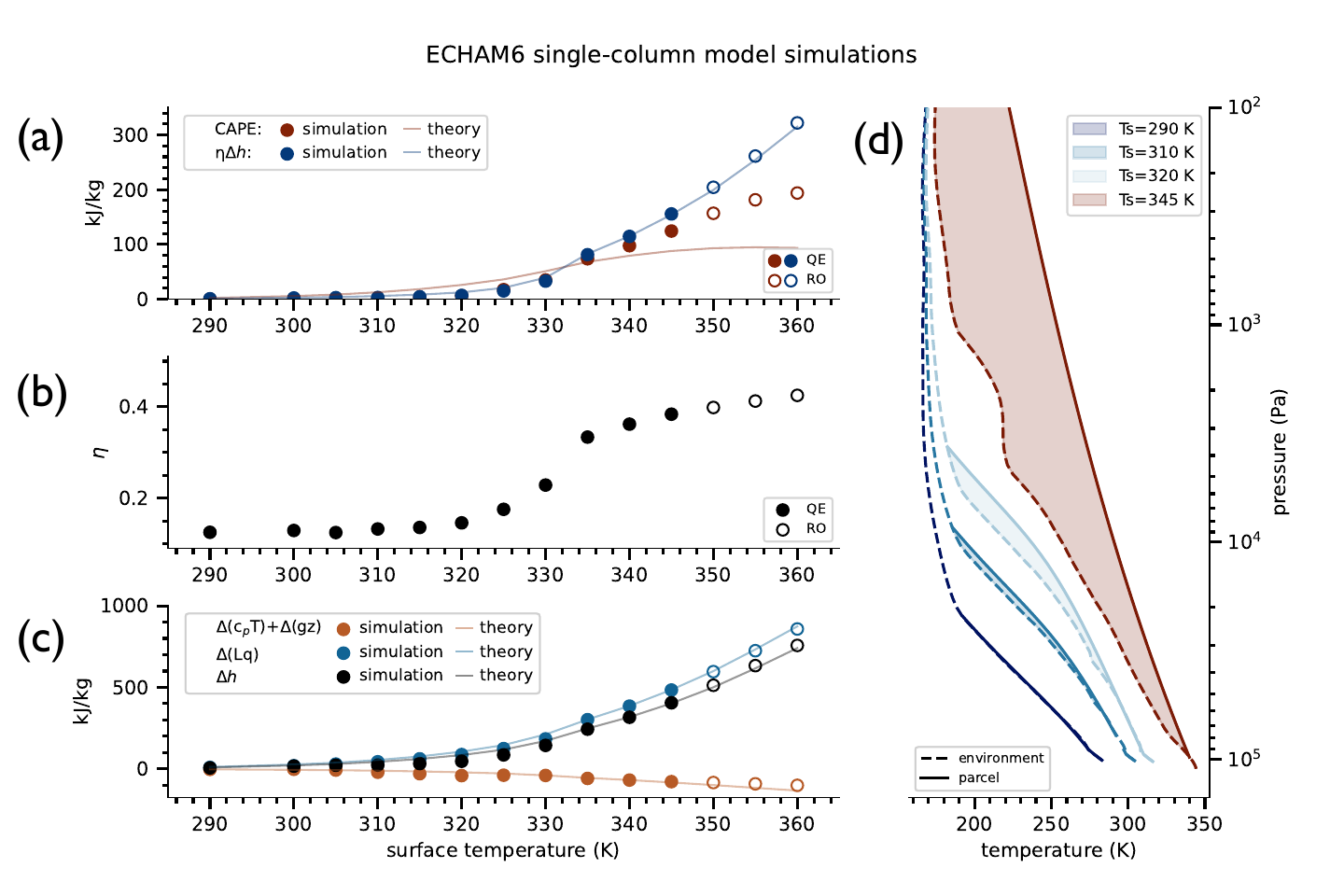}
  \caption{(a) Comparison of convective available potential energy (CAPE; Jkg$^{-1}$) and the vertical moist static energy difference ($\Delta h$; Jkg$^{-1}$) times the heat engine efficiency $\eta$ as a function of surface temperature. (b) Heat engine efficiency $\eta$ as a function of surface temperature. (c) Dry $\Delta(c_p T)$ and moist $\Delta (Lq)$ contribution to the moist static energy difference of upward and downward plumes across the LCL as a function of surface temperature. (d) Temperature difference of the environment and an adiabatically-lifted surface parcel in the simulation, given as a proxy for the steady-state CAPE. In (a)-(c), circular markers are diagnosed values from the base experiment with ECHAM6.} \label{fig:sam23b}
  \centering
\end{figure*}

Nonetheless, the existence of episodic precipitation in our one-dimensional simulations at high surface temperatures allows for a second test of the zero-buoyancy heat engine theory of convection. We determine the range of the convective parameters in the one-dimensional simulations over all surface temperatures to be $0.4\leq\text{PE}\leq0.8$ (Figure \ref{fig:conv-params}f), $0.55\leq\text{RH}\leq0.95$ (Figure \ref{fig:conv-params}g), and $0.3\leq a \leq 7.9$ (Figure \ref{fig:conv-params}h). As in Section \ref{sec:SW21}, we use PE and RH from the simulated QE states to diagnose an appropriate value for the bulk-plume parameter $a$ in the single-column model, yielding $\text{PE}=0.62$, $\text{RH}=0.81$, and $a=2.33$ (Figure \ref{fig:conv-params}f-h).

CAPE, $\eta$, and $\Delta  h$ from the simulations are shown as multi-decadal averages in Figure \ref{fig:sam23b}, alongside their predicted values from the theory ($\eta$ is exempt; see Section \ref{sec:SW21}). The heat transport by convection as measured by the vertical MSE difference increases from 9 kJ/kg to 860 kJ/kg between 290 K and 360 K. The work done by convection as quantified by CAPE ranges from 17 J/kg at 290 K to 194 kJ/kg at 360 K (Figure \ref{fig:sam23b}a). The heat engine efficiency increases from $\eta=$ 12\% to 42\% over the experimental surface temperature range (Figure \ref{fig:sam23b}b). The zero-buoyancy model tends to capture the overall trend of CAPE and $\Delta h$ in the simulations, though CAPE tends to be over-estimated and under-estimated at low and high surface temperatures, respectively. The theoretical fit to the simulated CAPE can be improved by relaxing the assumption of a single value for PE and $a$ (not shown). 

A comparative analysis of the metrics in the equilibrium condition yields several noteworthy findings. The first is that, while the theory reveals a peak in CAPE at intermediate surface temperatures \cite<>[]{Seeley2023}, we observe no clear peak in CAPE in the simulations because the environmental temperatures do not converge on the moist adiabat at high $T_s$ (Figure \ref{fig:sam23b}d); this can be understood as the result of the model having a large value of $a\sim6$ (Figure \ref{fig:conv-params}h). The second finding is that quasi-equilibrium is violated (CAPE $<\eta \Delta  h$) at intermediate surface temperatures in both the theory and the simulations (approximately 335 K in both cases; Figure \ref{fig:sam23b}a). This behavior arises due to the dependence of CAPE, $\Delta  h$, and $\eta$ on atmospheric moisture concentration. Specifically, the growth rate of CAPE, unlike $\Delta  h$, no longer conforms to the Clausius-Clapeyron rate at intermediate surface temperatures \cite{Romps2016,Seeley2023}. CAPE is curtailed by its dependence on the tropospheric heat capacity, which is itself affected by the ballooning concentration of water vapor \cite{Seeley2015,Seeley2016,Romps2016}. Meanwhile, $\eta$ increases with surface warming (Figure \ref{fig:sam23b}b). Below 320 K, this increase is primarily due to warming the ``hot" part of the system. Above 320 K, the radiative effects of water vapor cool the ``cold" part of system (i.e., $T_a$ decreases as the net radiating level ascends; Figures \ref{fig:rain-transition}a and \ref{fig:conv-params}e), which raises the efficiency as well. Third, RO-type precipitation is first observed in our simulations at  350 K (Figure \ref{fig:rain-transition}b). At this surface temperature, quasi-equilibrium is clearly violated (Figure \ref{fig:sam23b}a). However, the emergence temperature is 15 K higher than would be predicted under the strictest theoretical interpretation of QE breakdown. The online ECHAM6 calculation for CAPE uses buoyancy and lapse rate formulations that are behind the state-of-the-art (e.g., \ref{app:C}), which could also influence our offline interpretation of the single-column model results. Despite the caveats, the one-dimensional simulations seem to support the idea that violating the equilibrium condition (Equation \ref{eq:M=Msc}) leads to the emergence of RO convection. 

\section{Discussion}\label{sec:dis}

There have been several past studies of hothouse climates on Earth with one-dimensional and three-dimensional models with parameterized convection \cite{Wordsworth2013,W&T, Popp2015, Popp2016}. All of these studies found low-level temperature inversions in hothouse climates despite different model assumptions, however none of them reported episodic precipitation. The models of \citeA{W&T} and \citeA{Popp2015,Popp2016} employ different parameterizations, but the common elements included prognostic water ice, liquid, and vapor \cite{Rasch1998,Lohmann1996}, a bulk-plume convection scheme with quasi-equilibrium closure \cite{Zhang1995,Nordeng1994}, two-stream radiative transfer using the correlated-k method \cite{Wolf2013,Iacono2008}, parameterized ocean heat transport, and prognostic surface temperatures. In the three-dimensional studies, \citeA{W&T} included a seasonal cycle with modern ocean-land surface configuration, whereas \citeA{Popp2016} simulated a global aquaplanet with no seasonal cycle. \citeA{Seeley2021} were the first to report episodic precipitation in hothouse climates using a (regional) cloud-resolving model. For our single-column model simulations, we employed the same model as \citeA{Popp2015,Popp2016} in single-column mode with parameterized convection, but we followed the  experimental setup of \citeA{Seeley2021} with 10\% higher insolation than present-day, a mixed-layer ocean, and fixed sea surface temperatures. Under these experimental conditions, the single-column model produces episodic precipitation at high surface temperatures. 

It seems clear that relaxation-oscillator (RO) convection emerges in sufficiently warm and/or humid atmospheres. A novelty of this work is that RO convection is not only possible in  convection-resolving, three-dimensional simulations \cite{Seeley2021,Dagan2023}, but also in a single-column climate model with parameterized convection. That being said, the RO-type convection and precipitation that develops in our single-column simulations does not closely resemble the characteristics of RO convection in three-dimensional convection-resolving simulations \cite{Seeley2021}, which we address below.  

A second novelty of this work is the development of an even simpler model for the emergence of RO convection; we hypothesized that RO convection emerges in warm/humid climates due to a breakdown of QE convection, and developed a predictive theory based on a heat engine model for convection.  In QE, the rate of atmospheric radiative cooling can be related to a rate of work (Equation \ref{eq:HE}), which is proportional to the mass flux and CAPE. In sufficiently warm and humid atmospheres, the steady-state storage of CAPE is curtailed \cite{Seeley2023} as heating goes to the latent reservoir instead of increasing temperature and buoyancy \cite{Romps2016}. The change in CAPE with surface warming can be calculated using an analytical theory of convection in which entraining plumes are neutrally buoyant with respect to their environment \cite{Romps2016}, as cloudy regions are observed to be in Earth's tropics \cite{S&O}. Holding radiative cooling fixed, QE demands that the convective mass flux \emph{increase} with decreasing CAPE (Equation \ref{eq:M_HE}), as must occur at high temperatures where CAPE is decreasing. However, because the mass and energy budget of the convection must close in the sub-cloud layer, radiative cooling aloft is counterbalanced by convective heat transport, i.e. the MSE flux across the LCL.  It's intuitive that the MSE flux can grow rapidly as the surface temperature increases due to the exponential increase in saturation vapor pressure. Again holding radiative cooling fixed, this implies the sub-cloud mass flux must \emph{decrease} with increasing surface temperatures (Equation \ref{eq:M_sc}).  If the sub-cloud layer cannot supply the cloud layer with enough mass flux, there can be no QE state -- an inevitable consequence of the growing disparity between the increase in energy of the sub-cloud layer and a slower increase and/or decrease in CAPE.  This follows from the equilibrium condition (Equation \ref{eq:M=Msc}) of a convective heat engine, which requires (i) a statistical equivalence between vertical heat transport ($|M_{sc}| \times \Delta  h$; Equation \ref{eq:M_sc}) and radiative cooling ($F_a$; Equation \ref{eq:Ebudget}) and (ii) the conversion of surface heating into work ($|M| \times \text{CAPE}$; Equation \ref{eq:HE}) at a thermodynamic efficiency ($\eta$; Equation \ref{eq:eta}).

In addition to our theoretical arguments, we presented evidence from simulations with and without resolved convection in support of the idea that the RO mode of convection is preferred in conditions that violate the equilibrium condition. We emphasize that our analysis does not rule out the lower-tropospheric radiative heating hypothesis of \citeA{Seeley2021} as an explanation for RO emergence (see below); future work should compare these differing perspectives. Our analysis was first performed on convection-resolving model data from \citeA{Seeley2021}. We found that the equilibrium condition is violated around 320 K, which is consistent with the temperature of RO emergence in their simulations. We repeated the analysis on data from our one-dimensional model and found that $\eta \Delta  h$ first exceeds CAPE around 335 K, but that episodic precipitation emerges closer to 350 K. The RO state clearly emerges in conditions that violate QE, but the accuracy of the theoretical prediction is worse when applied to the single-column model. The heat engine perspective of convection suggests that the conditions that make the QE state energetically unsustainable -- in this case, curtailed growth in CAPE and sustained increases in atmospheric heat transport and opacity with surface warming -- causes the RO state to emerge. It is possible to obtain episodic convection in a one-dimensional climate model because modern parameterizations are able to represent the necessary physics in a single vertical dimension. This possibility is underscored by our usage of the bulk-plume equations of convection in the zero-buoyancy heat engine theory, which are, in turn, a simplified version of the \citeA{Nordeng1994} convection scheme in our one-dimensional climate model (Section \ref{sec:methods}). The existence of the QE-to-RO convective regime transition across the modeling hierarchy explored here lends confidence to the robustness of this transition, and we have demonstrated that important insights can be gained from the simpler end of the modeling hierarchy. 

Next, we discuss how our heat engine hypothesis connects with previous work on RO convection, starting with the discovery paper: \citeA{Seeley2021}. They found that the QE-to-RO transition coincides with the critical temperature at which the water vapor window closes, implying that LTRH is crucial for the emergence of RO states. However, subsequent studies by \citeA{Dagan2023} and \citeA{Song2024} challenged this hypothesis. Specifically, both studies demonstrated that while LTRH can be sufficient for the emergence of RO states in some cases, it is not a necessary condition. Meanwhile, \citeA{Song2024} investigated whether vertical contrasts in radiative cooling are more critical than the sign of the cooling rate in a specific atmospheric layer. They varied the magnitude of imposed cooling profiles in the lower and upper troposphere separately at a fixed surface temperature of 325 K. Their experiments revealed QE-to-RO transitions in response to substantial increases in upper-tropospheric radiative cooling or decreases in lower-tropospheric radiative cooling. In contrast, when radiative cooling was uniformly increased across the entire troposphere, no QE-to-RO transition occurred. These findings align with our theory, which posits that the efficiency $\eta$ is proportional to the difference between the inverse temperatures at which the engine absorbs ($1/T_s$) and emits ($1/T_a$) energy. Our simulations show that $T_a$ decreases with increasing surface temperature above 320 K, leading to an increase in heat engine efficiency with warming (Figure \ref{fig:sam23b}b), and this promotes QE breakdown. Given that $T_s$ is fixed in \citeA{Song2024}'s experiments, the emission-weighted atmospheric temperature, $T_a$, becomes critical. All else equal, adding a constant to the radiative cooling at all levels would exactly cancel out in the weighted average for $T_a$, which is consistent with the absence of a QE-to-RO transition in these runs in \citeA{Song2024}. Conversely, all else equal, enhancing upper-level cooling or suppressing lower-level cooling would decrease $T_a$, potentially leading to a violation of the equilibrium condition if $\Delta h$ and CAPE remain constant. To maintain quasi-equilibrium, what ``matters” more is not the rate at which heat is lost, but rather the temperature at which that heating or cooling occurs. This lesson is qualitatively consistent with the primary findings of \citeA{Song2024}.

The theory that we've developed tells us when steady convection \textit{must} break down, but not necessarily \textit{how} it happens. This limitation is intrinsic to equilibrium models, where there are no net forces. While it is reasonable to infer that steady convection ceases where quasi-equilibrium is incompatible with radiative-convective equilibrium, these constraints do not give insight into the physical forces acting on convective plumes at the QE-to-RO transition. Consequently, a common approach for determining why RO convection happens is to look for changes in CIN (Equation \ref{eq:cape(rho)}). In the non-equilibrium perspective, the presence of stable layers inhibits convection and delays convective triggering. There are several studies that have considered the effects of water vapor on inhibition \cite{Li2015,Seeley2021}. For example, water vapor plays an important role in modulating convective activity in Saturn’s atmosphere. The high molecular weight of water vapor compared to the non-condensing background gases suppresses convection until the slow cooling of the atmosphere makes it unstable. This mechanism has been put forward as an explanation for the intermittent, giant storms on Saturn \cite{Li2015}. In addition, the LTRH that is induced by water vapor in Earth's hothouse climates is important for generating very stable layers, including near-surface temperature inversions \cite{Wordsworth2013,W&T,Popp2015}. These inversions cap the boundary layer, decoupling it from the overlying atmosphere, and are slowly eroded by re-evaporation processes associated with descending virga. \citeA{Seeley2021}  demonstrated that imposing LTRH can tip the climate into the RO state. We speculate that this is also due to a breakdown of QE because LTRH likely increases $\eta$ by changing the effective absorbing and emitting temperatures, and reduces CAPE and/or introduces CIN. We leave it to future work to test this.

In our derivation of the zero-buoyancy heat engine model, we made a few notable assumptions, approximations, and omissions that introduce errors in the limit of a moisture-dominated atmosphere. Revisiting these assumptions is a necessary next step, but is beyond the scope of this paper. First, we assumed that a statistical balance exists between sources and sinks of entropy in Earth's climate system (Equation \ref{eq:Sbudget}), and furthermore that the dominant source is frictional dissipation and that the dominant sink is absorption and emission of radiation. This yields a prediction for the temperature of RO emergence, though at the cost of neglecting other potential sources of entropy generation. A simple balance between frictional dissipation and radiation is an excellent approximation in numerical simulations of dry convection \cite{Pauluis2002a}, and is also somewhat reasonable for moist convection on Earth. Irreversible mixing, phase changes, and hydrometeor drag are likely the dominant sources of entropy generation in Earth's current atmosphere \cite{Pauluis2002a,Pauluis2002b,Romps2008,Singh2022a}. Frictional dissipation (which includes hydrometeor drag) accounts for almost half of the total entropy generation in convection-resolving simulations \cite{Singh2016}. To improve our prediction of RO emergence, one could include the omitted processes in the entropy budget ($\Delta \dot S$; Equation \ref{eq:Sbudget-with-dS}) and repeat our analysis. Second, we neglected the potentially-important role of CIN in the vertical integral of buoyancy (Equation \ref{eq:cape}) and assumed that updrafts and downdrafts contribute equally to the buoyancy flux. Third, we made an important omission in the zero-buoyancy theory of convection: the virtual effect of the condensable gas. The virtual effect impacts the  magnitude of CAPE, particularly in condensable-rich atmospheres. Further exploration of these limitations is left to future work.

Our single-column model fails on important benchmarks set by convection-resolving models \cite{Seeley2021,Dagan2023}. First, the latter models agree that RO states on Earth should have a storm duration of several hours with an intervening period of several days. Our model produces storms with a duration of a few weeks that reoccur every few years. This could be related to the high PE of the single-column model simulations (Figure \ref{fig:conv-params}f), which we suspect is related to the non-zero precipitation during calm intervals in Figure \ref{fig:rain-transition}b; the convection scheme seems to have a minimum threshold for precipitation whether or not it would re-evaporate in lower layers due to a built-in assumption that there is no environmental re-evaporation of condensates above the base of the convective updrafts (see \ref{app:B}).  The model also produces very high values of CAPE, which can be understood by the large value of the inferred bulk plume parameter $a$ that prevents the environment from becoming moist adiabatic.  Second, our model transitions into the RO state at temperatures above 350 K, which is 30 K higher than in convection-resolving model simulations \cite{Seeley2021,Dagan2023,Song2024}.  This could be due to the large values of CAPE of the single-column model, which again is due to the large value of $a$; Figure \ref{fig:breakdown}a demonstrates that increasing CAPE values drives the QE breakdown to higher temperatures.  See \ref{app:B} for more details. The performance of the single-column model depends on the parameterizations and the assumptions made in developing them. These simple models are typically tuned to the modern climate of the planet they are meant to represent, and therefore are not guaranteed to represent reality in extreme scenarios. If the RO states in our single-column model are indeed driven by the same physics as in a convection-resolving model, then it is clear that our model suffers from poor realism, and we suspect the convective parameterization is to blame. It's conceivable that with some additional tuning, the single-column model could improve relative to the cloud-resolving models. 

\section{Implications for Titan}\label{sec:titan}

\begin{figure*}
\noindent\includegraphics[width=0.95\textwidth,angle=0]{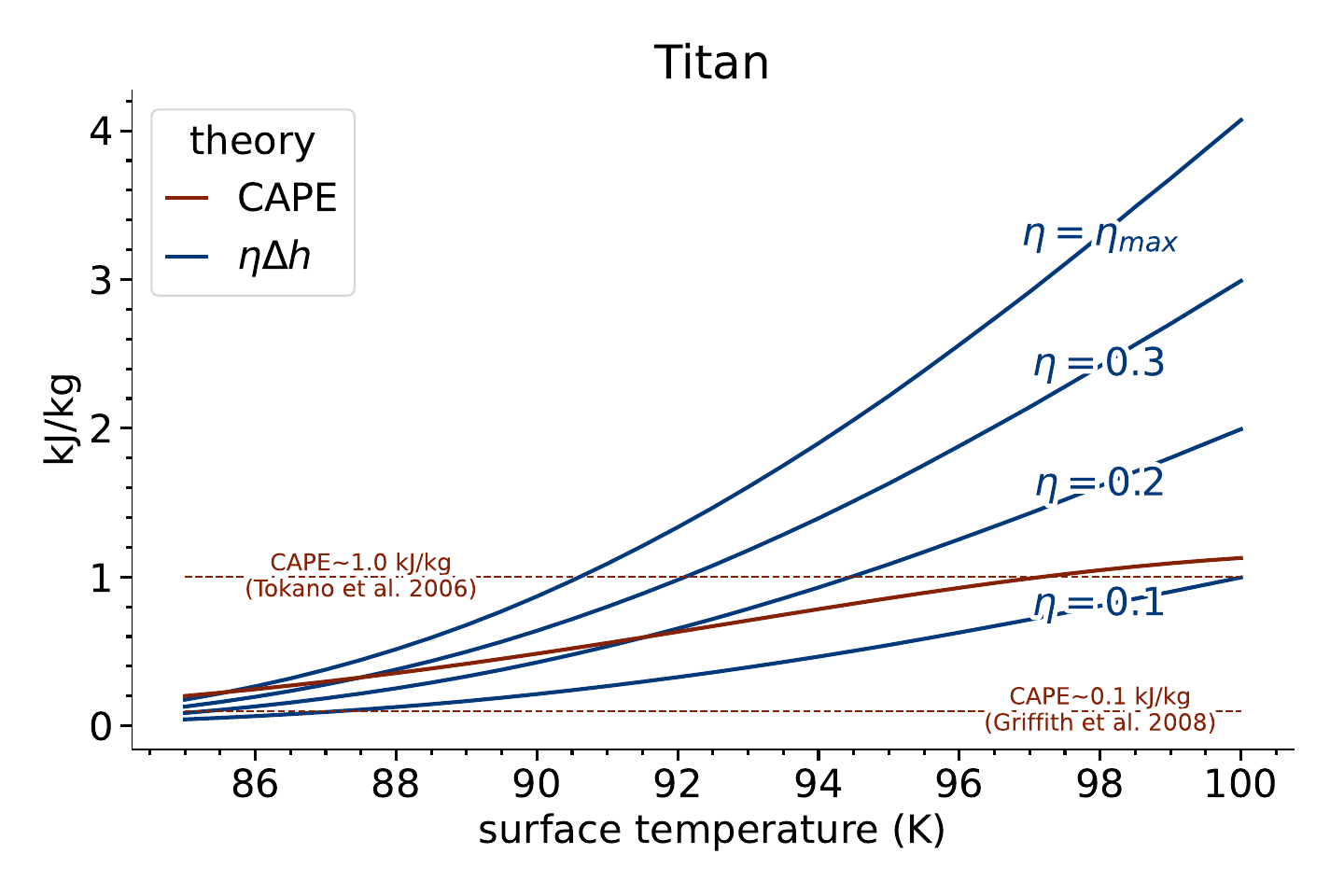}
  \caption{Theoretical CAPE (solid orange) and $\eta \Delta h$ (solid blue) for a Titan-like atmosphere from the zero-buoyancy heat engine model. In addition to the parameter choices for Titan in \ref{app:A}, we set $a=0.33$ and $\text{PE}=0.2$. Multiple labeled $\eta \Delta h$ curves show a hypothetical range of $\eta$ values between 0.1 and $\eta_{max}= (T_s - T_{trp})/T_{trp}$. Estimates of CAPE on Titan using the Huygens probe measurements disagree, with the estimates (dashed orange) ranging from $\sim$0.1 kJ/kg \cite{Griffith2008} to $\sim$1 kJ/kg \cite{Tokano2006}.} \label{fig:titan}
  \centering
\end{figure*}

While present-day Earth does not exhibit RO behavior (Figure \ref{fig:precip-earth-titan}a), it is clear that episodic storms of potentially great magnitude occur on present-day Titan \cite{Schneider2012,Turtle2011b,Faulk2017,Rafkin2022,Charnay2015,Mitchell2016}. Global simulations of Titan with realistic land surface hydrology by \citeA{Faulk2020} with the Titan Atmospheric Model \cite{Lora2015} depicted in Figure \ref{fig:precip-earth-titan}b resemble the one-dimensional (Figure \ref{fig:rain-transition}d) and three-dimensional \cite<Figure \ref{fig:precip-earth-titan}c;>[]{Seeley2021} simulated RO states on Earth, where brief periods of intense rainfall give way to extended periods of little to no rain. 

The intermittent cloud formation (and precipitation by inference) on Titan is presently understood to result from a combination of large-scale and convective-scale dynamics \cite{Mitchell2006}. Titan's cloud-forming regions undergo seasonal migration and concentrate near the summer pole due to thermally-direct global circulation promoting ascent in this area \cite{Turtle2011,Mitchell2016}. However, cloud formation is also driven by local convection related to seasonal changes in temperature and near-surface moisture \cite{Brown2002}. Observations support this, showing that high-latitude clouds often form in convectively unstable layers \cite{Griffith2000,Horst2017}. Given sufficient convective instability, storms can be triggered by poleward-propagating Rossby waves \cite{Battalio2021}. A regional convective outburst can also have a global impact \cite{Battalio2021,Turtle2011} through Rossby wave generation leading to outbreaks in other regions, such as over Titan’s low-latitude deserts \cite{Schaller2009,Turtle2011b}. 

The observed interval between large cloud systems on Titan is approximately 3-18 months \cite{Roe2012}, which might indicate the frequency of surface precipitation. To reproduce the frequency of these cloud-forming events, convective parameterizations in global climate models of Titan have to be tuned \cite{Battalio2022}. The Titan Atmospheric Model \cite{Lora2015}, from which the data in Figure \ref{fig:precip-earth-titan}b originates, uses a simple Betts-Miller convection scheme \cite{Frierson2007-BM}. In this scheme, convection is triggered when $\textrm{CAPE}-\textrm{CIN}>0$. Local conditions determine whether the triggered convection is precipitating or not. If the relative humidity is less (greater) than a preset reference value ($\textrm{RH}_{ref}$), then non-precipitating (precipitating) convection occurs. Using the value of $\textrm{RH}_{ref}$ most consistent with the observed frequency of Titan’s storms, the best-fit simulations from \citeA{Battalio2022} show a global-mean CAPE of 40 J/kg and a global-mean CIN of 50 J/kg. The authors attribute the existence of rare convective storms in their model to the fact that $\textrm{CAPE}-\textrm{CIN}<0$. In comparison, Earth's global-mean CAPE is 300 J/kg and global-mean CIN is 20 J/kg \cite{Riemann2009}. On the basis of the parameterization design, we would expect the same convection scheme in an Earth climate model to produce continuous precipitation. In the Titan study \cite{Battalio2022}, for example, the authors report a fundamental shift from episodic to continuous precipitation that occurs when $\textrm{CAPE}-\textrm{CIN}$ changes sign, consistent with this interpretation. Though some insight into the nature of Titan’s episodic precipitation can be gained from the parameterizations in general circulation models, regional or global convection-resolving simulations of Titan are needed.

The alignment of the climate of Titan and the hothouse Earth also merits further study. One example of this alignment is their total precipitable moisture: that is, the depth of liquid that would one would obtain by condensing out all atmospheric moisture into a uniform layer at the surface. Titan's atmosphere stores around $5$ m of precipitable methane \cite{Tokano2006}. Though the globally-averaged surface temperature on Earth is nearly 200 K warmer, the total precipitable moisture in Earth's atmosphere - just a few centimeters of water - is smaller because water on Earth has a volatility ten times less than methane on Titan \cite{Griffith2008,fsa23a}. Volatility is here defined as the saturation vapor pressure of a condensable at the typical temperature of the planetary surface. A consequence of the high temperatures in the hothouse climate is that the total precipitable water rises above 0.5 m by 330 K \cite{Seeley2021-data}, or to within an order of magnitude of the precipitable methane in Titan's atmosphere. Since the latent heat of condensation of water is about an order of magnitude larger than that for methane (\ref{app:A}), the latent energy reservoir (expressed in per unit area) in the hothouse climate is remarkably similar to Titan, which has important implications for moist convection.

Titan represents a unique laboratory for further testing of the zero-buoyancy heat engine theory of convection. To illustrate this, we carry out a back-of-the-envelope calculation of the equilibrium condition for Titan (Figure \ref{fig:titan}) with the parameter values in \ref{app:A}. The only direct measurement of CAPE in Titan's atmosphere comes from the Huygens probe, which landed at 10$^\circ$S where the near-surface relative humidity was $\sim50$\% \cite{Mitchell2016}. Estimates of CAPE at the time and location of the Huygen's probe descent vary, with values of 960 J/kg \cite{Tokano2006}, 60 J/kg \cite{Barth2007}, and 120 J/kg \cite{Griffith2008} reported. The disparity in these values supposedly stems from different formulations of the adiabatic lapse rate \cite{Griffith2008}. To the best of our knowledge, no direct estimation of the MSE profile at the Huygens landing site exists. Therefore, we rely on the spatially- and temporally-averaged vertical MSE difference from Titan Atmospheric Model simulations \cite<Figure \ref{fig:precip-earth-titan}b;>[]{Faulk2020}, yielding $\Delta h \approx 2500$ J/kg. MSE is well-mixed in the boundary layer and decreases with height from 5 km to 15 km, where the MSE minimum occurs (not shown). Since the surface temperature at Titan's equator is $\sim$95 K and the equator-to-pole temperature gradient is weak \cite<only 2-4 K;>[]{Jennings2011}, a typical surface temperature on Titan is likely 92 K. Configured to Titan-like conditions, the zero-buoyancy model predicts that CAPE$\sim$600 J/kg and $\Delta h \sim$3300 J/kg at 92 K (Figure \ref{fig:titan}). Theoretically, QE breakdown could occur on Titan if $\eta\geq20$\% at 92 K. However, a realistic estimate of $\eta$ on Titan is still needed for comparison with the zero-buoyancy heat engine model. Obtaining this would require convection-resolving simulations with realistic radiation, as $\eta$ depends on the unique radiative properties of Titan's atmosphere. On Earth, water vapor is the dominant greenhouse gas and sets the mean tropospheric cooling rate \cite{Held2000,JF20a}. Radiative processes are more complex on Titan, where the major source of infrared opacity is from collision-induced absorption of nitrogen, methane, and hydrogen \cite{McKay1991}. Titan's atmosphere is also strongly absorbing at solar wavelengths; 80\% of the incident flux is absorbed by the atmosphere and just 10\% at the surface \cite{Tomasko2008b}. 

An interesting question for future work is whether the dynamical similarity of Titan and the hothouse Earth can be linked to the breakdown of QE convection in both cases. The conceptual framework that we've proposed offers a robust point of comparison between planetary atmospheres based on the first and second laws of thermodynamics, which are system invariant. The theory is, moreover, agnostic of the composition of the working fluid and thus seems a promising framework to explore the dynamical similarity between Titan and the hothouse Earth. The theory could be extended to any planetary atmosphere with a condensing component and, because of this, it will find wide application in the solar system and beyond.

\appendix
\section{List of physical symbols, constants, and acronyms}\label{app:A}

\begin{sidewaystable}
\begin{tabular}{llll}
 %\hline
 Symbol & Definition & Earth-like & Titan-like \\
 \hline
 $a$ & Bulk-plume parameter & \\
 $A_u,A_d$ & Horizontal area of updrafts and downdrafts (m$^2$) & \\
 $B$ & Buoyancy (m s$^{-2}$) & \\
 $B_u,B_d$ & Buoyancy of updrafts and downdrafts (m s$^{-2}$) & \\
 $c$ & Condensation rate in convective updrafts (kg m$^{-3}$s$^{-1}$) & \\
$c_{va}$ & Isochoric specific heat of dry air (J kg$^{-1}$K$^{-1}$) & 719  & 707.2 \\
$c_{vv}$ & Isochoric specific heat of the condensable gas 
 (J kg$^{-1}$K$^{-1}$) & 1418 & 1707.4\\
$c_{vl}$ & Isochoric specific heat of the condensable liquid (J kg$^{-1}$K$^{-1}$) & 4119 & 3381.55 \\
$c_{vs}$ & Isochoric specific heat of the condensable solid (J kg$^{-1}$K$^{-1}$) &  &  \\
 $c_{p}$ & Isobaric specific heat of air  (J kg$^{-1}$K$^{-1}$) &  \\ 
 $c_{pa}=c_{va}+R_a$ & Isobaric specific heat of dry air  (J kg$^{-1}$K$^{-1}$) & 1006.04  & 1004. \\ 
 $c_{pv}= c_{vv}+R_v$ & Isobaric specific heat of the condensable gas(J kg$^{-1}$K$^{-1}$) & 1879 & 2225.68 \\
  $\Delta (c_p T) + \Delta (gz)$ & Dry static energy difference between the sub-cloud and cloud layer (J kg$^{-1}$) &  \\
 CAPE & Convective available potential energy (J kg$^{-1}$) & \\
  CIN & Convective inhibition (J kg$^{-1}$) & \\
 $\dot D$ & Rate of dissipation (W) & \\
 $d$ & Convective detrainment rate (kg m$^{-3}$s$^{-1}$) & \\ 
 $e$ & Convective entrainment rate (kg m$^{-3}$s$^{-1}$) & \\ 
 $e^*$ & Saturation vapor pressure of the condensable (Pa) & \\ 
 $f$ & Frictional force per unit mass (m s$^{-2}$) & \\  
  $\tilde F_a$ & Spatially-resolved net radiative flux from the atmosphere (W m$^{-2}$) & \\
$E_{0v}$ & Specific internal energy difference between condensable gas and liquid at $T_{trip}$ (J kg$^{-1}$) & $2.374 \times 10^6 $ & $4.9 \times 10^5$ \\
$E_{0s}$ & Specific internal energy difference between condensable liquid and solid at $T_{trip}$ (J kg$^{-1}$) & $3.34 \times 10^5$ & $5.86 \times 10^4$ \\
  $F_a$ & Net radiative flux from the atmosphere (W m$^{-2}$) & \\
  $F_{a,sc}$ & Net radiative flux from sub-cloud layer (W m$^{-2}$) & \\
  $F_{a,cl}$ & Net radiative flux from the cloud layer (W m$^{-2}$) & \\
  $\tilde F_s$ & Spatially-resolved net radiative flux at the surface (W m$^{-2}$) & \\
  $F_s$ & Net radiative flux at the surface (W m$^{-2}$) & \\
 $g$ & Gravitational acceleration (m s$^{-2}$) & 9.81 & 1.35 \\  
 $h$ or MSE & Moist static energy (J kg$^{-1}$) &   \\
  $h_{sc}$ & Moist static energy of the sub-cloud layer (J kg$^{-1}$) &   \\
  $h_{cl}$ & Moist static energy of the cloud layer (J kg$^{-1}$) &   \\
  $\Delta h$ & Moist static energy difference between the sub-cloud and cloud layer (J kg$^{-1}$) & \\
  $h^*$ & Saturation moist static energy (J kg$^{-1}$) &   \\
 $L$ & Latent heat of vaporization of the condensable (J kg$^{-1}$) & $2.5 \times 10^6$ & $5.5 \times 10^5$ \\
 $\Delta (Lq)$ & Latent energy difference between the sub-cloud and cloud layer (J kg$^{-1}$) &  \\
  LCL & Lifting condensation level &  \\
  LFC & Level of free convection &  \\
  LNB & Level of neutral buoyancy &  \\
 LTRH & Lower-tropospheric radiative heating & \\
  $M$ & Convective mass flux (kg m$^{-2}$s$^{-1}$) & \\
  $M\times B$ & Buoyancy flux (W m$^{-3}$) & \\
  $M_{sc}$ & Sub-cloud mass flux (kg m$^{-2}$s$^{-1}$) & \\
  $M_u,M_d$ & Mass flux of updrafts and downdrafts (kg m$^{-2}$s$^{-1}$) & \\
    $M_{u,sc}$ & Updraft mass flux in the sub-cloud layer (kg m$^{-2}$s$^{-1}$) & \\
    $M_{d,cl}$ & Downdraft mass flux in the cloud layer (kg m$^{-2}$s$^{-1}$) & \\
\end{tabular}
\end{sidewaystable}

\begin{sidewaystable}
\begin{tabular}{llll}
  PE & Precipitation efficiency & \\
 $p$ & Total atmospheric pressure (Pa) & \\ 
 $p_{trip}$ & Triple point pressure (Pa) & 611.65 & 11700.\\
 $P_s$ & Surface precipitation rate (kg m$^{-2}$s$^{-1}$) & \\
  $q$   & Specific humidity (kg kg$^{-1}$) & \\ 
  $q^*$ & Saturation specific humidity (kg kg$^{-1}$) & \\
  $\tilde Q_a$ & Spatially-resolved atmospheric cooling rate per unit mass (W kg$^{-1}$) & \\
  $Q_a$ & Atmospheric cooling rate (W) & \\
  $Q_s$ & Surface heating rate (W) & \\
 QE & Quasi-equilibrium & \\
 $R_a$ & Specific gas constant of dry air (J kg$^{-1}$K$^{-1}$) & 287.04  & 296.8\\
 $R_v$ & Specific gas constant of the condensable gas (J kg$^{-1}$K$^{-1}$) & 461 & 518.28  \\ 
 RH & Relative humidity & \\
 RO & Relaxation oscillator & \\
 $s$ & Specific entropy (J kg$^{-1}$K$^{-1}$) & \\
  $\dot s_d$ & Spatially-resolved specific entropy change due to frictional dissipation (W kg$^{-1}$K$^{-1}$) & \\
 $\dot s_{rad}$ & 
 Spatially-resolved specific entropy change due to radiative processes (W kg$^{-1}$K$^{-1}$ & \\
 $S$ & Entropy (J K$^{-1}$) & \\
$\dot S_d$ & Entropy change due to frictional dissipation (W K$^{-1}$) & \\
$\dot S_{irr}$ & Entropy change due to irreversible processes (W K$^{-1}$) & \\
$\dot S_{rad}$ & Entropy change due to radiative processes (W K$^{-1}$) & \\
$\Delta \dot S$ & Entropy change due to all irreversible processes except frictional dissipation (W K$^{-1}$) & \\
  SI & Vertically-integrated sink of condesable gas from phase changes (kg m$^{-2}$s$^{-1}$) & \\ 
    $T$ & Spatially-resolved atmospheric temperature (e.g., parcel/plume) (K) & \\
    $\overline T$ & Time- and horizontally-averaged temperature (K) & \\
  $T_a$ & Mean temperature at which atmospheric radiation is emitted (K) & \\
  $1/T_a$ & Mean inverse temperature at which atmospheric radiation is emitted (K$^{-1}$) & \\
  $T_d$ & Mean temperature at which frictional dissipation occurs (K) & \\
  $T_s$ & Surface temperature (K) & \\  
$T_{trp}$ & Tropopause temperature (K) & 200 & 71 \\  
$T_{trip}$ & Triple point temperature (K) & 273.16 & 90.68 \\
$v$ & Air velocity (m $s^{-1}$) & \\
$V$ & Volume ($m^{3}$) & \\
$\dot W$ & Rate of work (W) & \\
$\delta$ & Fractional detrainment efficiency (m$^{-1}$) & \\
$\varepsilon$ & Fractional entrainment efficiency (m$^{-1}$) & \\
$\eta$ & Efficiency of the convective heat engine & \\
$\eta_{max}$ & Maximum efficiency of the convective heat engine & \\
$\gamma$ & Moisture lapse rate (kg kg$^{-1}$m$^{-1}$) & \\
$\Gamma$ & Temperature lapse rate (K m$^{-1}$) & \\
$\rho$ & Spatially-resolved density of air (e.g., plume/parcel) (kg m$^{-3}$) & \\
$\overline \rho$ & Time- and horizontally-averaged density (kg m$^{-3}$) & \\
 \hline
%\caption{Parameters used to evaluate the zero-buoyancy heat engine model at Earth-like and Titan-like conditions. These are not necessarily the values used by ECHAM6 \citep{Stevens2013} or the Titan Atmospheric Model \citep[TAM -][]{Lora2015}.
%\label{table:parameters}}
\label{table:parameters}
\end{tabular}
\end{sidewaystable}

\section{Diagnosis of convective parameters: results and interpretation}\label{app:B}
\begin{figure*}[b]
  \centering
  \includegraphics[width=\textwidth,angle=0]{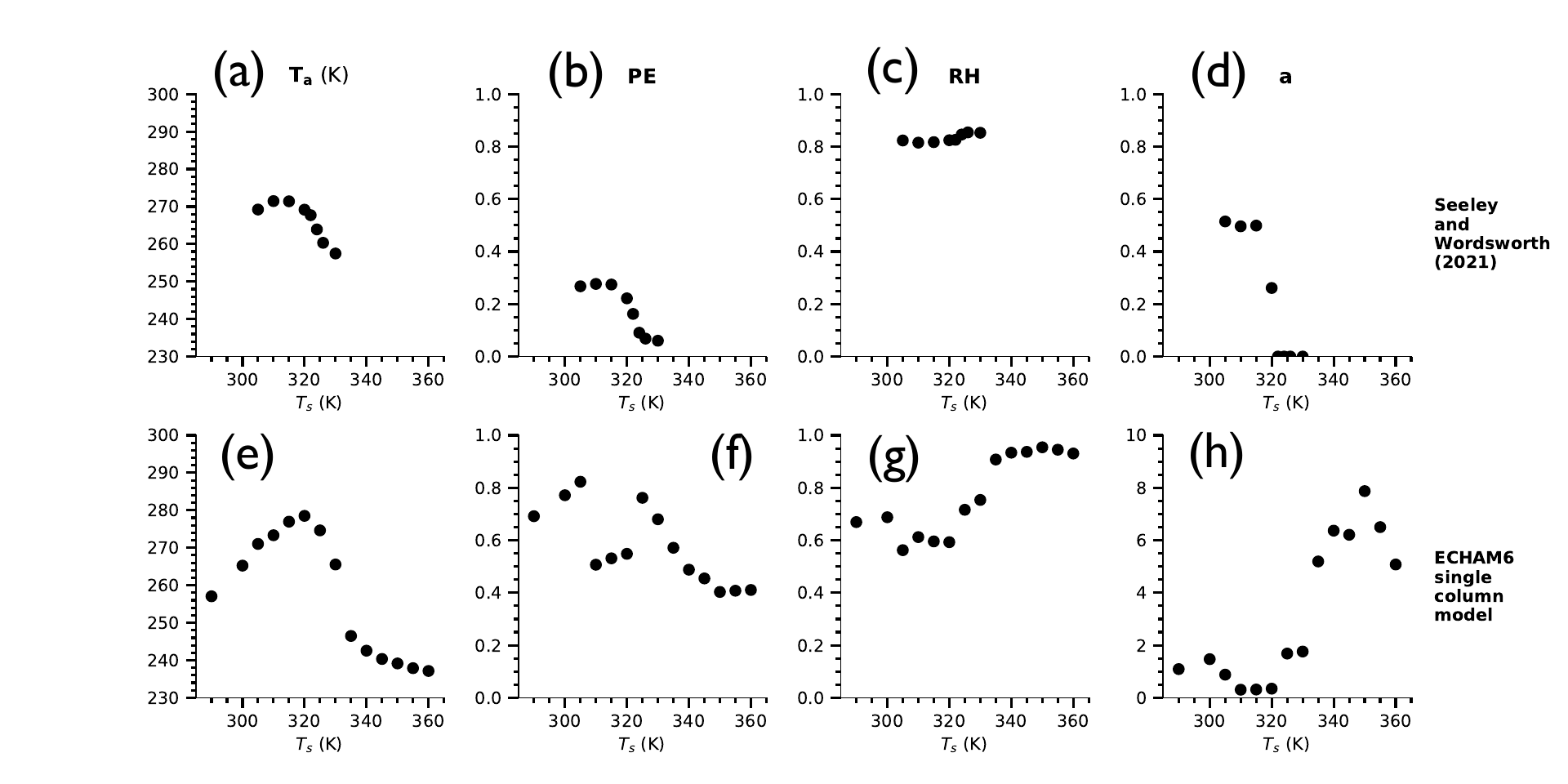}
  \caption{As a function of surface temperature, (a,e) mean temperature at which radiation is emitted, $T_a$, (b,f) precipitation efficiency PE, (c,g) mean tropospheric relative humidity RH, and (d,h) bulk-plume parameter $a$ from simulations with (a--d) a convection-resolving model \cite{Seeley2021} and (e--h) the ECHAM6 single-column model.} \label{fig:conv-params}
\end{figure*}

\begin{figure*}
  \centering
  \includegraphics[width=\textwidth,angle=0]{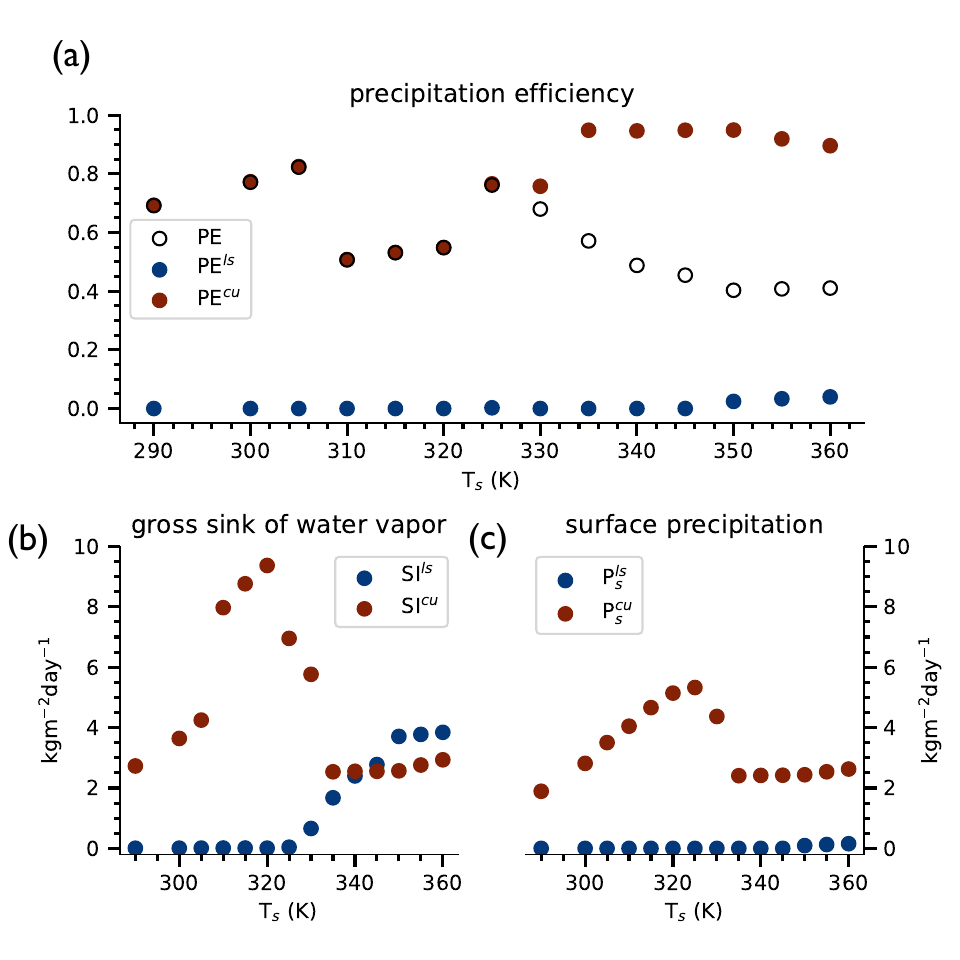}
  \caption{Diagnosed quantities from the ECHAM6 single-column model. (a) PE compared to the precipitation efficiency of the large-scale scheme PE$^{ls}$ and the convection scheme PE$^{u}$. For visual clarity, PE is an unfilled marker. (b) Vertically-integrated gross sink of water vapor due to phase changes in the large-scale scheme SI$^{ls}$ and the convection scheme SI$^{u}$. (c) Surface precipitation from the large-scale scheme P$_s^{ls}$ and the convection scheme P$_s^{u}$.} 
  \label{fig:PE-echam}
\end{figure*}

\begin{figure*}[b]
  \centering
  \includegraphics[width=\textwidth,angle=0]{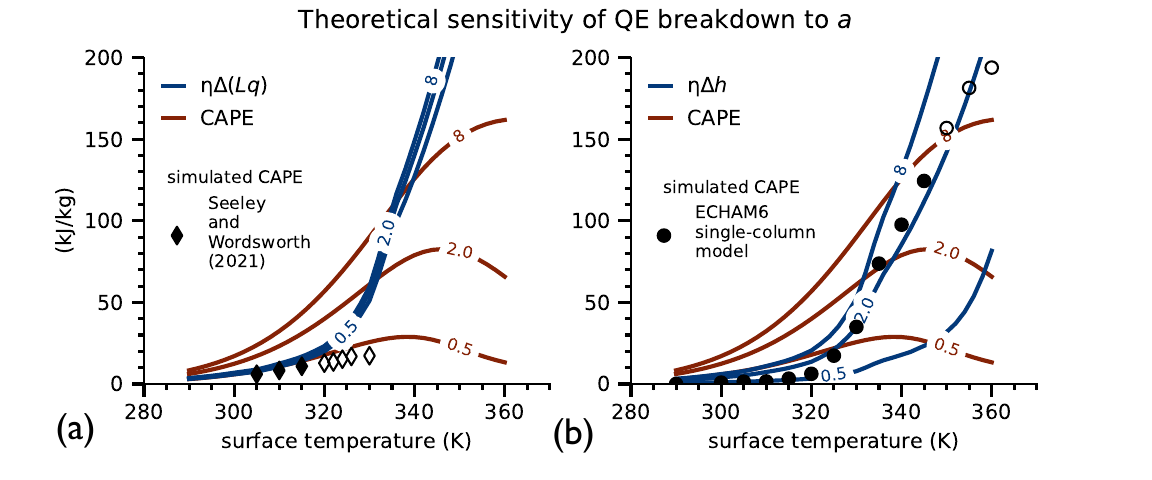}
  \caption{Theoretical sensitivity of the QE breakdown hypothesis to varying the bulk-plume parameter $a$. The (solid) theoretical profiles are labeled by values of $a$ between 0.5 and 8. We initialize the parcel at an assumed pressure of 1 bar, the temperature of the surface, and PE=25\%. In (a), we estimate the vertical heat transport as the vertical latent energy difference $\Delta (Lq)$ only. The simulated CAPE from \citeA{Seeley2021} is plotted for reference (symbols). In (b), we estimate the vertical heat transport as the vertical moist static energy difference $\Delta h$. The simulated CAPE from the single-column model is plotted for reference. As in the main text, filled markers indicate QE states and unfilled markers indicate RO states.} 
  \label{fig:a-sens}
\end{figure*}

Figure \ref{fig:conv-params} shows the mean temperature at which radiation is emitted from the atmosphere $T_a$, the precipitation efficiency PE, the mean tropospheric relative humidity RH, and the bulk-plume parameter $a$, as diagnosed from the convection-resolving simulations of \citeA{Seeley2021} and the ECHAM6 single-column model. Below, we discuss important trends and inter-model differences. 

First, we consider the inter-model differences in PE (Figure \ref{fig:conv-params}b,f). Besides its relevance to the hydrological cycle, we care about PE because it allows us to diagnose $a$. The trend in PE with surface temperature in the convection-resolving model is not reproduced by the single-column model. In the convection-resolving model, PE takes values between 5-30\%. The PE in the single-column model is substantially larger at all surface temperatures (up to a factor of 10). To understand why, we diagnose the precipitation efficiencies of the large-scale and convection schemes in the single-column model separately. The convective and large-scale precipitation efficiencies are $\text{PE}^{u}=P^{u}_s/\text{SI}^{u}$ and $\text{PE}^{ls}=P^{ls}_s/\text{SI}^{ls}$ (see also Section \ref{sec:PE-echam}). Note that these quantities are not strictly additive (i.e., they do not sum to PE). These separate metrics for precipitation efficiency, gross condensation, and surface precipitation are shown in Figure \ref{fig:PE-echam}. Figure \ref{fig:PE-echam}b reveals that there are two condensation regimes. For surface temperatures below 325 K, the majority of gross condensation in the atmosphere is convective in origin (Figure \ref{fig:PE-echam}b). For this reason, $\text{PE}$ is biased towards $\text{PE}^{u}$ below 325 K (Figure \ref{fig:PE-echam}a). Above 325 K, the fraction of gross condensation in updrafts decreases while the fraction of gross condensation in the large-scale environment increases, until the two are roughly equal in magnitude at 335 K (Figure \ref{fig:PE-echam}b). Large-scale and convective sources of precipitation are displayed in Figure \ref{fig:PE-echam}c, demonstrating that the latter dominates the total precipitation at every surface temperature. The bias of surface precipitation to the convection scheme is a well-documented behavior of general circulation models \cite{Chen2021}. Total surface precipitation is curtailed above 320 K (Figure \ref{fig:PE-echam}c) as a result of the water vapor window closing and shortwave absorption increasing, which is a robust phenomenon in hothouse climates \cite{Liu2024}. Comparing the separate sources of gross condensation and surface precipitation, it is clear that the convection scheme is significantly more efficient than the large-scale scheme. Consistent with this interpretation, $\text{PE}^{u}$ ranges from 50-100\% and $\text{PE}^{ls}$ is at most a few percent (Figure \ref{fig:PE-echam}a). Lastly, there is a gradual decrease in PE between 325 to 350 K (Figure \ref{fig:PE-echam}a). Figure \ref{fig:PE-echam}b,c demonstrates that this decrease is related to the increasing fraction of gross condensation originating in the large-scale scheme, which biases the PE towards $\text{PE}^{ls}$. 

Why is the convection scheme so much more efficient than the large-scale scheme in the single-column model? The answer lies in the parameterization design. In the convection scheme, a fraction of the condensates produced in updrafts are converted into convective precipitation. The rest is detrained into the environment, and is then converted into stratiform clouds or large-scale precipitation by the large-scale scheme. The large-scale precipitation is re-evaporated as it descends through the sub-saturated environment, but the convective precipitation is \textit{assumed} to fall through the saturated updraft (this precludes re-evaporation of convective precipitation in the middle troposphere) and exits from the updraft base near the surface. This ``insulation" of convective precipitation from environmental conditions within the plumes supports the persistent near-surface drizzle. In contrast, large-scale precipitation is exposed to sub-saturated environmental conditions during descent, which could explain why the overwhelming majority of gross condensates are re-evaporated before reaching the surface (Figure \ref{fig:PE-echam}c). The difference in PE between the schemes is likely due to these assumptions. The convection-resolving model simulations tell us that the real PE should be closer to that of the large-scale scheme at high surface temperatures (Figures \ref{fig:conv-params}f and \ref{fig:PE-echam}a). It might be possible to tune our model to have a lower PE at high temperatures in accordance with the convection-resolving simulations. We leave this as a future task, as the bias in the partitioning of rainwater in climate models is widespread \cite{Chen2021} and beyond the scope of our work. 

Using the PE (Figure \ref{fig:conv-params}b,f) and mean tropospheric RH from the simulations (Figure \ref{fig:conv-params}c,g), we are able to diagnose the bulk-plume parameter in both models. $a$ is found to be less than 0.5 in the convection-resolving simulations on account of the high RH and low PE (Figure \ref{fig:conv-params}b,c). In the context of the zero-buoyancy theory (Section \ref{sec:ZBA}), this means that convective plumes and environmental air are moderately coupled. $a$ is substantially larger in the single-column model; it ranges from 0.3-1.5 below 325 K and from 1.5-8 above 325 K (Figure \ref{fig:conv-params}h). There is a significant increase in $a$ above 325 K (Figure \ref{fig:conv-params}h) related to the decline in PE, rather than the increase in RH (the latter would act to reduce $a$). Figure \ref{fig:a-sens} shows the effect of varying $a$ between 0.5-8 on the theoretical CAPE. In general, larger values of $a$ support more steady-state storage of CAPE (e.g., Figure \ref{fig:a-sens}a). Low values of $a$ give a better fit to the convection-resolving model simulations (Figure \ref{fig:a-sens}a), as might be expected from Figure \ref{fig:conv-params}d. Conversely, $a=8$ yields values for CAPE that approach those observed at high surface temperatures in the single-column model (Figure \ref{fig:a-sens}b). 

Interestingly, the theoretical temperature of QE breakdown is sensitive to the value of $a$ (Figure \ref{fig:a-sens}b). We are interested in this sensitivity as a potential explanation for the 30 K difference in the surface temperature of RO emergence between the models. As you increase $a$, the temperature of QE breakdown decreases (Figure \ref{fig:a-sens}b). Why is this? The dry static energy has a strong dependence on $a$ in the theory because of the location of the minimum in tropospheric MSE. In the zero-buoyancy theory, decreasing $a$ reduces the lapse rate, which causes the MSE minimum to rise to a higher elevation. This means that the dry static energy difference becomes more negative (and therefore reduces $\Delta h$) with decreasing $a$. That $a$ is larger in the single-column model does not seem to explain why the RO state emerges at a lower surface temperature in the convection-resolving model.

However, the theoretical prediction for QE breakdown is sensitive to the method of estimating vertical heat transport. In this study, we chose to use a simple difference of the MSE between the surface and the tropospheric minimum value. Consider the scenario in Figure \ref{fig:a-sens}a where $\Delta h$ is approximated by the latent energy $\Delta (Lq)$. In this case where we have neglected the vertical dry static energy difference, the temperature of QE breakdown is instead found to \textit{increase} with increasing $a$. This result is more consistent with our intuition, and leads us to suspect that there is a better way to estimate the vertical heat transport. Future work should therefore interrogate our assumptions about the vertical heat transport in an atmosphere with zero buoyancy. We assumed that upward air records the MSE near the surface and downward air records the minimum in MSE, but is that consistent with zero buoyancy? The environmental temperature and moisture profile in the zero-buoyancy model is a reflection of the mutual interaction (entrainment, detrainment, re-evaporation) between ascending and descending air. In this process, air parcels exchanged across the same level have the same dry static energy, and differ only in their moisture content. This might mean that the appropriate measure of vertical heat transport in the zero-buoyancy atmosphere is not $\Delta h$, but rather $\Delta (Lq)$ or some variant thereof.

\section{How to calculate CAPE in the simulations}\label{app:C}
This appendix summarizes our process of calculating CAPE in the simulations \cite{Romps2008,Romps2015,Marquet2016,Emanuel1994}. To ensure consistency in our analysis, we apply the same parcel methods to the data from the convection-resolving model of \citeA{Seeley2021} and our single-column model. The values of thermodynamic constants for Earth-like and Titan-like atmospheres are given in \ref{app:A}.

Moist air is defined to be a mixture containing dry air and a condensable in various phases. The mass fraction (i.e. specific humidity) is represented by the symbol $q_x$. $x$ is a generic subscript referring either to dry air $a$, condensable gas $v$, condensable liquid $l$, or condensable solid $s$. The mixing ratio is represented by the symbol $r_x$, and is related to the specific humidity by $r_x = q_x/q_a$. The density of dry air is $\rho_a = p_a/(R_a T)$. The partial pressure of dry air in the parcel is a function of the total pressure $p$ (given by the simulation) and the mixing ratio of the condensable gas $r_v$:
\begin{equation}
p_a = p (1+r_v/\epsilon)^{-1},
\end{equation}
where $\epsilon=R_a/R_v$. When the specific volume of liquid and solid phases of the condensable are neglected, the density of the moist air parcel is \cite{Emanuel1994} 
\begin{equation}
\label{eq:rhomoist}
\rho = \frac{p}{R_a T} \frac{1+r_t}{1+r_v/\epsilon}, 
\end{equation}
where $r_t = r_v + r_l + r_s$ is the total mixing ratio of the condensable. For simplicity, we will ignore the solid phase ($r_s,q_s=0$) in our parcel calculations and assume that, regardless of temperature, the condensable exists only in the gas and/or the liquid phase; however, we will retain variables associated with ice in subsequent equations for reader clarity. The saturation vapor pressure over liquid is \cite{Romps2008,Romps2015}
\begin{equation}
e^{*,l} = p_{trip} \Bigg(\frac{T}{T_{trip}} \Bigg)^{(c_{pv}-c_{vl})/R_v} \times \exp \Bigg[ \frac{E_{0v}-(c_{vv}-c_{vl})T_{trip}}{R_v} \Bigg(\frac{1}{T_{trip}} - \frac{1}{T} \Bigg)   \Bigg],
\end{equation}
where $p_{trip}$ and $T_{trip}$ are the triple point pressure and temperature, $c_{pv}$ is the isobaric specific heat of condensable gas (J kg$^{-1}$ K$^{-1}$), $c_{vv} $  and $c_{vl}$ are the isochoric specific heats of condensable gas and condensable liquid (J kg$^{-1}$ K$^{-1}$), and $E_{0v}$ is the difference in specific internal energy between condensable gas and condensable liquid at the triple point temperature \cite<J kg$^{-1}$;>[]{Romps2015}. The saturation specific humidity over liquid is then given by
\begin{equation}
q^{*,l} = \frac{\rho^*_v}{\rho} = \frac{R_a}{R_v} \frac{e^{*,l}}{p} \frac{1+r_v/\epsilon}{1+r_t}
\end{equation}

To determine the lapse rate of an ``adiabatic parcel", we invoke the conservation of the sum of MSE and CAPE \cite{Marquet2016,Romps2015}: that is, MSE + CAPE. Here, ``adiabatic" refers to any parcel that is raised without exchanging heat or mass with its surroundings, irrespective of the assumptions about moisture removal in the lifting method. For pseudo-adiabatic ascent, we remove all moisture after each discrete lifting step. For reversible ascent, the total moisture is conserved over the entire ascent. Taking the vertical derivative of the aforementioned conserved quantity and using the definition of CAPE, 
\begin{equation}
    \label{eq:dzMSE}
    \frac{\partial \text{MSE}}{\partial z} = -\frac{\partial }{\partial z} \int_{\textrm{LFC}}^{z} B(z') \: dz' = -B(z)
\end{equation}
where 
\begin{equation}
\label{eq:MSE}
\text{MSE} = [q_ac_{pa}+(q_v + q_l + q_s)c_{vl}](T-T_{trip}) + q_v L_c - q_s L_f + gz.
\end{equation}
$c_{pa}$ is the isobaric specific heat of dry air (J kg$^{-1}$ K$^{-1}$), $L_c = E_{0v} + R_v T + (c_{vv}-c_{vl})(T-T_{trip})$ is the latent heat of condensation (J kg$^{-1}$), and $L_f = E_{0s} + (c_{vl}-c_{vs})(T-T_{trip})$ is the latent heat of fusion (J kg$^{-1}$). $c_{vs}$ is the isochoric specific heat of condensable solid (J kg$^{-1}$ K$^{-1}$), and $E_{0s}$ is the difference in specific internal energy between condensable liquid and condensable solid at the triple point temperature \cite<J kg$^{-1}$;>[]{Romps2015}.

For detailed instructions on how to obtain the temperature and density profile of an adiabatic parcel with the same pressure as the local environment using conservation of MSE$+$CAPE, we refer the reader to \citeA{Romps2015}. However, we do offer a short summary below. MSE and $B$ at height $z$ depend on $T$ and $q_x$. The MSE at $z + \Delta z$ follows from Equation \ref{eq:dzMSE}. Solving for the parcel temperature at $z + \Delta z$ from the MSE (Equation \ref{eq:MSE}) requires a root solver. The reason is that MSE is a function of both temperature and the condensable mass fraction, where the partitioning between vapor, liquid, and solid phases is itself temperature-dependent. The solution constraints are that the vapor phase must remain pegged to the saturation value above the LCL and total moisture must be conserved during each discrete lifting step. When the level of neutral buoyancy is reached, the buoyancy $B(z)$ can be integrated upward from the LFC to yield CAPE. 

We have thus far detailed the parcel method for reversible ascent, where the total moisture in the parcel is conserved. We now discuss how to approach lifting scenarios where total moisture is not conserved, such as pseudo-adiabatic ascent.

\subsection{Ascent with condensate removal}
An example of irreversible ascent is where the condensed liquid or solid in the parcel is removed in part or in whole (i.e., pseudo-adiabatic). This process usually carries away a small amount of mass. The changing mass of the parcel produces a change in the specific humidity, which is accounted for as follows. The initial mass of the parcel is $m^i_{tot} = m_a^i + m_v^i + m_l^i + m_s^i$. The superscripts $i$ and $f$ are used to track the initial and final state of the parcel. For simplicity, we only detail the treatment of liquid removal, but note that ice removal would proceed analogously. Condensate removal is parameterized as exponential decay following \citeA{Seeley2023}:
\begin{equation}
    \label{eq:cremoval}
    \frac{\partial q_l}{\partial z} = - q_l/L,
\end{equation}
where $L$ is a characteristic length scale. Suppose that we remove a mass of condensable liquid $\Delta m_l<0$ as prescribed by Equation \ref{eq:cremoval}. The final and initial masses are related by $m_a^i=m_a^f$, $m_v^i = m_v^f$, and $m_l^f=m_l^i + \Delta m_l$, where we have neglected ice. The final mass fraction for dry gas and condensable gas are
\begin{equation}
    q_a^f = \frac{m_a}{m_{tot}^i + \Delta m_l} = \frac{q_a^i}{1 + \Delta q_l} \: \: \text{and} \: \: q_v^f = \frac{m_v}{m_{tot}^i + \Delta m_l} = \frac{q_v^i}{1 + \Delta q_l},
\end{equation}
where in the second step we divided through by $m_{tot}^i$. The liquid removal step increases the mass fraction of dry gas and vapor in the parcel by a factor of $(1+\Delta q_l)^{-1}$. The final mass fraction for condensable liquid is
\begin{equation}
    q_l^f = \frac{m_l^f}{m_{tot}^f} = \frac{m_l^i+\Delta m_l}{m_{tot}^i + \Delta m_l} = \frac{q_l^i + \Delta q_l}{1+\Delta q_l},
\end{equation}
where we have accounted for the changing total mass in addition to the total mass of condensable liquid. This method can be used to evaluate lifting processes from fully reversible ($L\rightarrow \infty$) to pseudo-adiabatic ($L\rightarrow 0$). The final mass fractions are used to update the parcel buoyancy and MSE before the next discrete lifting step.

%%%%%%%%%%%%%%%%%%%%%%%%%%%%%%%%%%%%%%%%%%%%%%%
% Optional Glossary, Notation or Acronym section goes here:
%
% Glossary is only allowed in Reviews of Geophysics
%  \begin{glossary}
%  \term{Term}
%   Term Definition here
%  \term{Term}
%   Term Definition here
%  \term{Term}
%   Term Definition here
%  \end{glossary}

%%%%%%%%%%%%%%%%%%%%%%%%%%%%%%%%%%%%%%%%%%%%%%%
% Acronyms
%% NOTE that acronyms in the final published version will be spelled out when used in figure captions.
%   \begin{acronyms}
%   \acro{Acronym}
%   Definition here
%   \acro{EMOS}
%   Ensemble model output statistics
%   \acro{ECMWF}
%   Centre for Medium-Range Weather Forecasts
%   \end{acronyms}

%%%%%%%%%%%%%%%%%%%%%%%%%%%%%%%%%%%%%%%%%%%%%%%
% Notation
%   \begin{notation}
%   \notation{$a+b$} Notation Definition here
%   \notation{$e=mc^2$}
%   Equation in German-born physicist Albert Einstein's theory of special
%  relativity that showed that the increased relativistic mass ($m$) of a
%  body comes from the energy of motion of the body—that is, its kinetic
%  energy ($E$)—divided by the speed of light squared ($c^2$).
%   \end{notation}

%%%%%%%%%%%%%%%%%%%%%%%%%%%%%%%%%%%%%%%%%%%%%%%
%
% DATA SECTION and ACKNOWLEDGMENTS
%
%%%%%%%%%%%%%%%%%%%%%%%%%%%%%%%%%%%%%%%%%%%%%%%

\section*{Open Research Section}
The data from simulations conducted with the modified version \cite{Popp2015} of the atmospheric general circulation model ECHAM6 \cite{Stevens2013} in single column mode is available at \citeA{sam24a-data}. Data from the NASA MERRA-2 climate reanalysis is attributed to \citeA{MERRA2-data}. Fixed surface temperature simulation data of the hothouse Earth from the convection-resolving model DAM \cite{Romps2008} is attributed to \citeA{Seeley2021-data}. Global simulation data of Titan from the general circulation model TAM \cite{Lora2015} is attributed to \citeA{Faulk2020-data}.

%\section*{As Applicable – Inclusion in Global Research Statement}

\acknowledgments
This work was supported by NSF Grant 1912673 and an Early-Career Fellowship from the Center for Diverse Leadership in Science at UCLA. The authors thank Jake Seeley, Bowen Fan, and Namrah Habib for helpful discussions, Robin Wordsworth and an anonymous reviewer for valuable comments on an early draft of this manuscript, and Jake Seeley for recommending the use of a root finder in our adiabat solver. We are grateful to 3 anonymous reviewers for suggestions that improved the final manuscript.

%%%%%%%%%%%%%%%%%%%%%%%%%%%%%%%%%%%%%%%%%%%%%%%
% REFERENCES and BIBLIOGRAPHY
%
% \bibliography{<name of your .bib file>} don't specify the file extension
% don't specify bibliographystyle
%
%%%%%%%%%%%%%%%%%%%%%%%%%%%%%%%%%%%%%%%%%%%%%%%

\bibliography{agu}

%Reference citation instructions and examples:
%
% Please use ONLY \cite and \citeA for reference citations.
% \cite for parenthetical references
% ...as shown in recent studies (Simpson et al., 2019)
% \citeA for in-text citations
% ...Simpson et al. (2019) have shown...
%
%
%...as shown by \citeA{jskilby}.
%...as shown by \citeA{lewin76}, \citeA{carson86}, \citeA{bartoldy02}, and \citeA{rinaldi03}.
%...has been shown \citeA{jskilbye}.
%...has been shown \citeA{lewin76,carson86,bartoldy02,rinaldi03}.
%... \cite <i.e.>[]{lewin76,carson86,bartoldy02,rinaldi03}.
%...has been shown by \cite <e.g.,>[and others]{lewin76}.
%
% apacite uses < > for prenotes and [ ] for postnotes
% DO NOT use other cite commands (e.g., \citet, \citep, \citeyear, \nocite, \citealp, etc.).
%

\end{document}